\documentclass[english,11pt,oneside]{article}

\usepackage{amsmath}
\usepackage{amssymb}
\usepackage{cite}
\usepackage{graphicx}
\usepackage{bbm}
\usepackage{color}
\usepackage{url}
\usepackage{authblk}
\usepackage{mathtools}
\usepackage{blindtext}
\usepackage{titlesec}
\usepackage{hyperref}

\newcommand{\be}{\begin{equation}}  
\newcommand{\ee}{\end{equation}}

\newcommand{\vev}[1]{\langle #1 \rangle}

\newcommand{\SO}[1]{\ensuremath{\mathrm{SO}(#1)}}
\newcommand{\SU}[1]{\ensuremath{\mathrm{SU}(#1)}}
\newcommand{\U}[1]{\ensuremath{\mathrm{U}(#1)}}
\newcommand{\into}{\ensuremath{\,\rightarrow\,}}
\newcommand{\tr}{\operatorname{tr}}

\newcommand{\MP}{\ensuremath{M_\mathrm{P}}}
\newcommand{\GN}{\ensuremath{G_\mathrm{\scriptscriptstyle{N}}}}

\definecolor{deepfuchsia}{rgb}{0.76, 0.33, 0.76}

\begin{document}

\title{\bf Hybrid inflation and gravitational waves \\
from accidentally light scalars}

\author[1]{Felix Br\"ummer}
\author[1]{Giacomo Ferrante}
\author[2]{Michele Frigerio}
\affil[1]{ \emph{Laboratoire Univers et Particules de Montpellier (LUPM), University of Montpellier and CNRS, Montpellier, France}}
\affil[2]{ \emph{Laboratoire Charles Coulomb (L2C), University of Montpellier and CNRS, Montpellier, France}}

\date{\today}
\maketitle
\begin{abstract}
 \noindent 
We construct a hybrid-inflation model where the inflaton potential is  generated radiatively, as gauge symmetries guarantee it to be accidentally flat at tree level. The model can be regarded as a small-field version of Natural Inflation, with inflation ending when the mass of a second scalar, the waterfall field, turns tachyonic. 
This provides a minimal, robust realization of hybrid inflation, which
predicts specific correlations among Cosmic Microwave Background observables.
Tachyonic preheating leads to the production of gravitational waves which,
for a low inflationary scale, might be detected by upcoming experiments.  Simple variations of the model can give rise to topological defects, such as unstable domain walls. 
Their dynamics produces a stochastic gravitational-wave background, which
can be compatible with the recent detection  by pulsar timing arrays. 
\end{abstract}

\newpage
\tableofcontents

\section{Introduction}

Field-theoretical models of cosmic inflation can be roughly divided into two classes. In large-field models, the field excursion of the inflaton is transplanckian, while in small-field models it remains below the Planck scale. When allowing for large field excursions, very simple shapes of the potential can give rise to inflation (although many are now excluded by increasingly accurate Cosmic Microwave Background (CMB) measurements \cite{Planck:2018vyg, Planck:2018jri}). However, it is generally dubious if large-field models can be trustable effective field theories, without being embedded in a complete quantum gravity framework. Small-field models, on the other hand, generally require more involved field-theoretic constructions. In particular, some mechanism should be included to protect the inflaton potential from radiative corrections. Once this is achieved, effects from transplanckian physics can be argued to be under control in the effective field theory. The only necessary fine-tuning left is that of the cosmological constant after inflation, and potentially of the initial conditions leading to an inflationary phase.

In small-field models, the flatness of the inflaton potential could be ensured, for example, by (spontaneously broken) supersymmetry, or by an (approximate) shift symmetry if the inflaton is a pseudo Nambu-Goldstone boson (pNGB). In the present work, we propose to employ a novel mechanism to this end: The inflaton is an accidentally massless scalar field, i.e., a scalar whose potential is constrained by symmetry to be flat up to loop corrections and higher-dimensional operators \cite{Brummer:2023znr}. Similarly to the pNGB case, a small slope is generated radiatively. 

At first sight, our inflaton potential is very similar to that of Natural Inflation (NI), where the inflaton is realized as a $\U{1}$ pNGB \cite{Freese:1990rb}. Minimal NI, however, is a large-field model, and, moreover, is excluded by Planck data \cite{Planck:2018jri}. Small-field NI needs additional fields responsible for ending inflation, and such models of ``hybrid NI'' turn out to be difficult to construct, since the couplings between the additional fields and the inflaton should not spoil the flatness of the potential (although this can be achieved with suitable discrete symmetries \cite{Ross:2009hg, Ross:2010fg}).

Our model, by contrast, is simple and economical; in the minimal case, the symmetry is simply a gauged $\SO{3}$ times a $\mathbb{Z}_2$ parity. Indeed, models with accidentally light scalars \cite{Brummer:2023znr} lend themselves particularly well to hybrid inflation model-building. This is because they feature an almost-flat inflaton direction by construction, yet the other scalar fields couple to the inflaton at the tree level, which allows inflation to naturally end by a second field becoming tachyonic.

We will compute the CMB observables as a function of the model parameters, and study their correlations. We will also present naturalness arguments to identify the preferred region of parameters, and derive the corresponding predictions.

While naturalness would prefer an inflationary
scale not too far from the Planck scale, 
lower-scale scenarios are more interesting for the potential observation of gravitational wave (GW) signals.
In hybrid models such as ours, inflation ends because a steep tachyonic ``waterfall'' direction appears in field space \cite{Linde:1993cn}, which leads to a transition from the inflationary slow-roll regime to a phase of tachyonic preheating \cite{Felder:2000hj, Felder:2001kt}. In this phase, GWs can be produced, and may be within the reach of future detectors, if the inflationary scale is low.
In addition, spontaneous symmetry breaking (SSB) in  the inflaton-waterfall potential may lead to the production of topological defects, which also source GWs.
We will present slightly extended versions of the minimal model, which produce cosmic strings or domain walls after the end of inflation. These scenarios are phenomenologically interesting only if the inflationary scale is sufficiently low, as, in this case, the annihilations of topological defects produce GWs which can be within 
experimental reach.

Primordial GW sources can potentially explain the recent observation of a stochastic background by the pulsar-timing-array (PTA) collaborations NANOGrav \cite{NANOGrav:2023gor}, EPTA and InPTA \cite{EPTA:2023fyk}, PPTA \cite{Reardon:2023gzh}, and CPTA \cite{Xu:2023wog}.  
We will compute the various GW signals in our scenario, study their correlations with the CMB observables, and compare them with present constraints from  LIGO-Virgo-KAGRA \cite{KAGRA:2021kbb} and  NANOGrav \cite{NANOGrav:2023hvm}, as well as with the expected sensitivity of future GW detectors.

Both large-field and small-field models are the subject of several swampland conjectures, which aim to characterize the set of effective field theories that are supposedly incompatible with general principles of quantum gravity; see Ref.~\cite{Palti:2019pca} for a review. Specifically, models with transplanckian field excursions are concerned by the Distance Conjecture \cite{Ooguri:2006in}. This conjecture states roughly that, at large distances in field space, an infinite tower of fields becomes light, invalidating the effective field theory. On the other hand, slow-roll potentials with subplanckian field excursions and positive vacuum energy are claimed to be ruled out by various (arguably more speculative) de Sitter conjectures \cite{Obied:2018sgi,Garg:2018reu,Andriot:2018wzk}. While these conjectures are certainly interesting, none of them is firmly established. For this work, we, therefore, assume that single-field slow-roll inflation is not per se in the swampland, and that small-field models, in particular, can be viable effective field theories of inflation.

\section{Hybrid inflation with an accidentally light scalar}\label{sec:AccInf}

Models with accidentally massless scalar fields were recently constructed in Ref.~\cite{Brummer:2023znr}, where the focus was on models with a single scalar in an irreducible representation and no ad-hoc discrete symmetries. To construct a model of hybrid inflation, however, we need scalar fields in at least two different multiplets, one of which will contain the inflaton and the other the waterfall field. While our model will share many features with the ones presented in Ref.~\cite{Brummer:2023znr}, the details of its field content and symmetry structure differ. 

The model is defined by a $G=\SO{3}$ symmetry with a real scalar $\phi$ transforming in the ${\bf 5}$ of $\SO{3}$, and a real scalar $\chi$ transforming in the ${\bf 3}$.
We also impose a $\mathbb{Z}_2$ symmetry under which $\phi$ is odd. The most general renormalizable potential is
\be\label{V5}
 V=-\frac{1}{2}\mu_\phi^2\phi^2-\frac{1}{2}\mu_\chi^2\chi^2+\frac{\lambda_\phi}{4}(\phi^2)^2+\frac{\lambda_\chi}{4}(\chi^2)^2+\frac{\varepsilon}{4}\phi^2\chi^2+
 \frac{\zeta}{4}\,T^a_{AC} T^b_{CB}\phi_{A} \phi_B \chi^{a}\chi^b\,.
\ee
Here, $T^a_{BC}$ are the $\SO{3}$ generators in the ${\bf 5}$-plet representation ($a=1,2,3$ and $A,B,C=1\ldots 5$). We refer the interested reader to Appendix \ref{sec:AppendixDetails} for a detailed derivation of Eq.~\eqref{V5}, and a comprehensive analysis of the vacuum structure of this potential.

The continuous symmetry of $V$ is strictly $G=\SO{3}$: There is no larger custodial-type symmetry present. More precisely, the symmetry of the kinetic terms for $\phi_A$ and $\chi_a$ is $\SO{8}$, which is explicitly broken to $\SO{5}_\phi\times\SO{3}_\chi$ by the mass terms and the quartic couplings $\lambda_{\phi,\chi}$ and $\epsilon$. It is non-trivial to show that a single remaining coupling, $\zeta$, breaks
explicitly ten generators of this group, reducing the symmetry to the ``diagonal'' $\SO{3}$, which acts on both $\phi$ and $\chi$. The absence of a larger symmetry implies that, after spontaneous symmetry breaking, there will be at most three Nambu-Goldstone bosons (NGBs); therefore, the appearance of additional, accidentally massless scalars is a non-trivial phenomenon.

It is instructive to consider first the case $\mu_\phi^2>0$ and $\mu_\chi^2<0$ with all quartic couplings positive. 
This parameter choice does not lead to hybrid inflation, but it is useful in order to understand how a slow-roll direction emerges. The potential is minimized at $\vev{\chi}=0$ and $|\vev{\phi}|=v$, where $v^2=\mu_\phi^2/\lambda_\phi$. 
Therefore, the vacuum manifold is a four-sphere, with $G$ completely broken at generic points. Three of the four flat directions are Nambu-Goldstone directions corresponding to the broken $G$ generators. If $G$ is gauged, the associated NGBs are absorbed by the gauge bosons. The fourth flat direction corresponds to an accidentally (tree level-) massless scalar field. It is not associated to any symmetry of the scalar potential and as such is neither an NGB nor, strictly speaking, a pNGB. The flat direction will be lifted by loop corrections through both the scalar quartic coupling  $\zeta$ and the gauge coupling $g$. In the limit of small $\zeta$, it can be understood as a genuine pNGB direction, as the scalar potential acquires a larger symmetry for $\zeta\to 0$.

At a special point along the flat direction, corresponding to a null eigenvector of one of the $G$ generators, an O$(2)=\SO{2}\rtimes\mathbb{Z}_2$ subgroup of $G$ is preserved, where the symbol $\rtimes$ denotes a semidirect product. At this point, there are only two NGBs, while not one but two scalars remain accidentally massless at the tree level; the second one is the scalar which is absorbed by the $\SO{2}$ gauge boson outside the special point. This point turns out to be a minimum of the one-loop effective potential.

For concreteness, consider an explicit parametrization of the tree-level flat direction by writing the isospin-2 field $\phi$ as a $3\times 3$ traceless symmetric matrix $\Phi=\phi_A \lambda_A$. Here, $\lambda_A$ are the the symmetric Gell-Mann matrices $\lambda _{1,3,4,6,8}$. By an $\SO{3}$ transformation, $\Phi$ can be diagonalized, such that its vacuum expectation value (VEV) becomes
\be\label{eq:alpha}
\langle\Phi\rangle=v\left(\lambda _3\sin\alpha+\lambda _8\cos\alpha\right)\,,\qquad v^2=\mu_\phi^2/\lambda_\phi\,.
\ee
The mass of the radial mode in $\Phi$ is
\be
m_\rho^2=2\lambda_\phi v^2
\ee
and the $\chi$ masses are
\be\label{eq:mchi}
m_{\chi^{1,2}}^2=-\tilde\mu_\chi^2+2\,v^2\zeta\,\sin^2\left(\alpha\pm\frac{\pi}{3}\right)\,,\qquad
m_{\chi^3}^2=-\tilde\mu_\chi^2+2\,\zeta v^2\sin^2\alpha
\ee
where we have defined
\be\label{eq:muchieff}
\tilde\mu_\chi^2=\mu_\chi^2-\frac{\varepsilon}{2}v^2\,.
\ee

Since, for the time being, we chose $\mu_\chi^2<0$ and $\epsilon$ and $\zeta$ positive, we have $\tilde\mu_\chi^2<0$ and so all masses are positive. The $\alpha$-dependence of the masses is a crucial feature of this model which will eventually allow us to realize hybrid inflation. 
As $G$ is gauged, the gauge bosons are also massive at the tree level.
The vacuum manifold of physically inequivalent points is parametrized by $\alpha\in[0,\,\pi/6]$, since points related by $\alpha\into -\alpha$ and $\alpha\into\alpha+\pi/3$ are identified under $G$. At $\alpha=0$, there is a symmetry-enhanced point with a massless $\SO{2}$ gauge field and one additional massless scalar field (which is otherwise absorbed by the third massive gauge boson).

Let us now consider the effect of one-loop corrections to the potential. The procedure to calculate the one-loop effective potential is detailed in Appendix \ref{sec:AppendixDetails}. 
When including one-loop corrections, the vacuum degeneracy is lifted, the physical vacuum is at the point $\alpha=0$, and the two tree-level-massless scalars become massive. The only massless state is the gauge boson of the unbroken $\SO{2}$. However, there remains a one-loop mass hierarchy between the six heavy scalar modes having obtained their masses at the tree level, and the two accidentally light ones with one-loop masses. 
We emphasize that the latter are calculable and finite as functions of $\mu_\phi^2$ and $\mu_\chi^2$, and, therefore, they are quadratically sensitive to quantum corrections at the cutoff scale only at two loops (since contributions to the mass parameters themselves are quadratically divergent at one loop), much as in Little Higgs models (for a review of the latter, see, e.g., Ref.~\cite{Schmaltz:2005ky}).

Focusing on the one-loop effective potential along the tree-level flat direction, one finds that it is given by
\footnote{See Appendix \ref{sec:AppendixDetails}
for the renormalisation prescriptions
employed to determine $V_{\rm eff}$,
and for the proof that it
can be expanded as  $\sum_n c_n \cos(na/f)$.}
\be\label{eq:CosPot}
V_{\rm eff}(a)=V_0-M^4\cos\frac{a}{f}+(\text{higher harmonics})~,
\ee
where $a=\alpha v$ is the canonically normalized ``accident'' field parametrizing the tree-level flat direction, $f=v/6$, and we have arbitrarily renormalized the vacuum energy to be $V_0-M^4$.
Moreover, we find
\be\label{eq:M4}
M^4 =\frac{1}{160\,\pi^2}\Bigg[
9\,g^4\,v^4
+2\,\frac{\tilde\mu_\chi^{10}}{\zeta^3\,v^6}\Bigg(F\left(\frac{\zeta v^2}{\tilde\mu_\chi^2}\right)
-T_{F,5}\left(\frac{\zeta v^2}{\tilde\mu_\chi^2}\right)\Bigg)\Bigg]~,
\ee
where $g$ is the $\SO{3}$ gauge coupling, $F(x)=(1-2\,x)^{5/2}$, and $T_{F,5}(x)=1-5\,x+\frac{15}{2}x^2-\frac{5}{2}x^3-\frac{5}{8}x^4-\frac{3}{8}x^5$ is the Taylor polynomial of $F(x)$ of degree 5.
Higher harmonics in the effective potential are suppressed relative to the first harmonic by small numerical factors.\footnote{For example, the coefficient of the second harmonic $\cos(2a/f)$ induced by $\zeta$ can be shown to always be smaller than its counterpart of Eq.~\eqref{eq:M4} by a factor $\gtrsim 10$ in the region of field space of interest. Likewise, the second harmonic induced by $g$ is always suppressed with respect to the first.} They will be neglected in the following.

The shape of the one-loop potential, therefore, closely resembles what one would obtain if $a$ were the pNGB of an approximate global $\U{1}$ symmetry. Identifying $a$ with the inflaton and setting $V_0=M^4$, one obtains a realization of the famous NI model \cite{Freese:1990rb}. However, vanilla NI is excluded by Planck data \cite{Planck:2018jri}. Simple modifications of the minimal NI model, such as a non-minimal coupling to gravity \cite{Ferreira:2018nav, Reyimuaji:2020goi,Salvio:2023cry} or allowing for some exotic phase of reheating \cite{Stein:2021uge}, can improve the fit, but in the preferred parameter region one finds $f\gtrsim \MP$.\footnote{We designate by $\MP$ the reduced Planck mass: $\MP \equiv (8\pi \GN)^{-1/2} = 2.4\times10^{18}\,{\rm GeV}$.} As already discussed, transplanckian field excursions, apart from being constrained by the Swampland Distance Conjecture, imply that the effects of higher-dimensional operators are not under control (although models have been constructed which address this issue \cite{Arkani-Hamed:2003xts,Kaplan:2003aj}).
To overcome these problems, it is desirable to identify a variant of the model where the field excursion is sub-Planckian, i.e., a small-field model of inflation.

In a small-field version of NI, the inflationary potential should be dominated by the constant term $V_0$ in the vacuum energy. This implies that slow-roll is possible for $f<\MP$, but it also implies that slow-roll never ends on its own, because the slow-roll parameters never grow to become ${\cal O}(1)$, and that the vacuum energy density $V_0$ somehow needs to be converted into radiation eventually. To this end, one may extend the model into one of hybrid inflation \cite{Linde:1993cn}: add a second scalar field, the waterfall field, which is heavy during inflation but becomes tachyonic for some value of the inflaton, triggering the end of inflation and the start of preheating. 

Models of hybrid inflation with pNGBs as inflatons were constructed, for example, in Refs.~\cite{Garcia-Bellido:1996mdl,Kaplan:2003aj}. In these models, the approximate symmetry protecting the inflaton is explicitly broken by the couplings to the waterfall field. These must be chosen ad hoc such as to preserve the flatness of the potential. Related models were subsequently built \cite{Ross:2009hg, Ross:2010fg}, where it was shown that the required structure of the potential can be protected by non-abelian discrete symmetries. By contrast, we will now show that our model with an accidentally light scalar can furnish a natural model of hybrid inflation with the simple potential of Eq.~\eqref{V5}.

So far, the role of the field $\chi$ was solely to provide a source for loop corrections to the effective potential which lift the flatness of the inflaton potential. We will now identify $\chi$ with the waterfall field by choosing the model parameters such that the true vacuum of the model is no longer at $\vev{\chi}=0$ and $\vev{\phi}\neq 0$ but rather at $\vev{\phi}=0$ and $\vev{\chi}\neq 0$. Our previous discussion of the vacuum structure applies as long as the effective $\chi$ masses, given in Eq.~\eqref{eq:mchi}, are all positive, regardless of the sign of the mass parameter $\tilde\mu_\chi^2$ in Eq.~\eqref{eq:muchieff}. Even for $\tilde\mu_\chi^2>0$, the effective potential is still given by Eqs.~\eqref{eq:CosPot} and \eqref{eq:M4} upon replacing $M^4$ by its real part. It is only when a tree-level mass becomes tachyonic that our computation of $V_{\rm eff}$ can no longer be trusted. 

Now choosing parameters such that $\tilde\mu_\chi^2$ satisfies
\be\label{eq:mubound}
0 < \tilde\mu_\chi^2 <\frac{\zeta}{2} v^2\,,
\ee
one observes that $m^2_{\chi^3}$ is positive for $\alpha=\pi/6$, negative for $\alpha=0$, and crosses zero at
\be\label{eq:alphac}
\alpha_c=\arcsin\sqrt{\frac{\tilde\mu_\chi^2}{2\zeta\,v^2}}\,,
\ee
while all other masses remain strictly positive for $\alpha\geq\alpha_c$.
This can be achieved by choosing $\mu_\chi^2>0$ and suitable (positive) values for the quartics. 

\begin{figure*}[bt]
\begin{center}
      \includegraphics[width=13 cm]{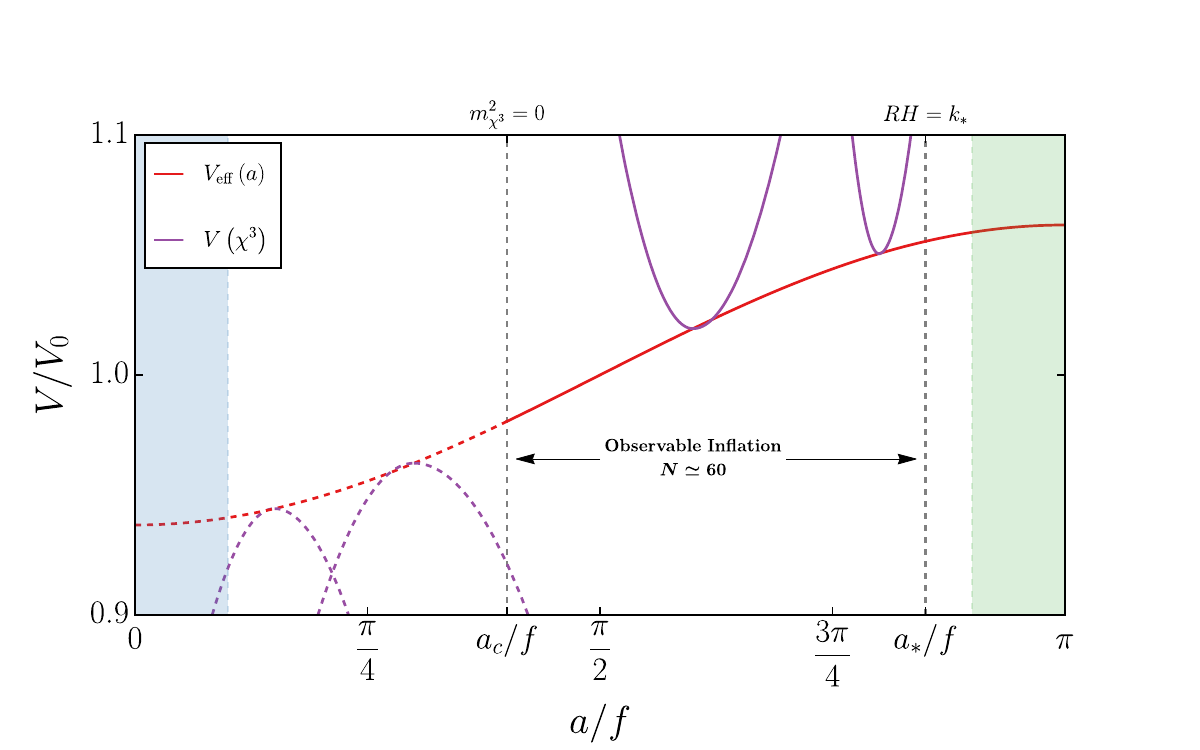}
    \caption{Pictorial representation of the inflationary potential, normalised to the inflationary scale $V_0$. Close to $a = \pi f$, where inflation begins, the potential along the $\chi^3$ direction (purple lines) has positive and large curvature, stabilizing the inflationary trajectory (red line). At the $a=a_*$ the pivot scale crosses the Hubble horizon, $R H = k_*$ (with $R$ the scale factor). As the inflaton slow-rolls down its potential, $m_{\chi^3}^2$ decreases until it becomes $0$ at the critical point $a_c$, where inflation ends. For $a<a_c$, the effective potential cannot be trusted, as the accidentally flat direction is no longer a minimum of the tree-level potential. A value of $a_*/f$ ($a_c/f$) in the green (blue) region would correspond to a tuning $\gtrsim 10\%$ on the initial (final) point of observable inflation, and, therefore, it is theoretically disfavoured.}
\label{fig:InflaSketch} 
\end{center}
\end{figure*}

In this regime, the inflaton $a$ starts rolling close to $\alpha=\pi/6$ or, equivalently, $a=\pi f$, where all scalars except $a$ have positive tree-level masses. In particular, $\chi$ is stabilized at $\chi=0$. For $\alpha>\alpha_c$, the one-loop effective potential Eq.~\eqref{eq:CosPot} can be trusted, and provides a small slope which the inflaton rolls down. As the inflaton crosses the point $\alpha_c$, the tree-level vacuum will no longer be given by $\chi=0$ and Eq.~\eqref{eq:alpha}, since a tree-level tachyon develops in the $\chi^3$ direction. Slow-roll ends as the tachyonic $\chi$ mass grows large, and $\chi$ modes are being excited. Eventually, the potential energy of the inflaton is converted into energy in the $\chi$ sector. In Fig.~\ref{fig:InflaSketch}, we provide a sketch of the potential along the $a$ (inflaton) and $\chi^3$ (waterfall) directions, for the sake of illustration.
Reheating should then proceed via some coupling, such as a Higgs portal coupling, between $\chi$ and the Standard Model; in this work, we will not explore reheating in detail.
It is easily checked that, for suitable values of the couplings, the true vacuum of the model is located at
\be
\chi^3=\sqrt{\frac{\mu^2_\chi}{\lambda_\chi}} \equiv v_\chi\,,\qquad
\text{all other fields} = 0
\ee
up to $G$-transformations. The residual symmetry is $\SO{2}\subset G$, and all fields except for the unbroken gauge boson pick up tree-level masses. Since the  true vacuum should be Minkowski up to a tiny cosmological constant, we can identify
\be\label{eq:V0}
V_0\simeq\frac{\mu_\chi^4}{4\,\lambda_\chi}-\frac{\mu_\phi^4}{4\,\lambda_\phi}\,,
\ee
which is the (tree-level) potential energy difference between the inflationary field configuration and the true vacuum. In the following, we will assume that the  second term in Eq.~\eqref{eq:V0}  is not only subdominant, but even negligible.

\section{Inflation parameters and CMB observables}\label{sec:CMB}

As inflation proceeds along the tree-level flat direction, the waterfall field $\chi$ is stabilized at $\chi=0$ by its effective mass; hence, the model qualifies as a single-field model. The evolution of scalar and tensor perturbations is then dictated by the potential in Eq.~\eqref{eq:CosPot}, with the slow-roll parameters defined by
\be
    \epsilon \equiv \frac{\MP^2}{2}\left(\frac{V_{\rm eff}'}{V_{\rm eff}}\right)^2\,,\qquad \eta \equiv \MP^2 \frac{V_{\rm eff}''}{V_{\rm eff}}\,,\qquad \xi \equiv \MP^4\frac{V_{\rm eff}'V_{\rm eff}'''}{V_{\rm eff}^2}\,,\qquad \sigma \equiv \MP^6 \frac{\left(V'_{\rm eff}\right)^2 V''''_{\rm eff}}{V_{\rm eff}^3}\,,
\ee
where primes denote derivatives with respect to the inflaton field $a$. 

At the leading order in the slow-roll parameters, the tensor-to-scalar ratio $r$, the amplitude of the curvature power spectrum $A_s$, the spectral index of the curvature power spectrum $n_s$, the running of the spectral index $n_r$, and the running of the running $n_{rr}$ can be computed, respectively, as
\be
\begin{aligned}
    &r = 16 \epsilon_*\,, \qquad A_s=\frac{H_*^2}{8\pi^2\MP^2}\frac{1}{\epsilon_*}\,, \qquad  n_s = 1 + 2\eta_* - 6\epsilon_*\,,\\ 
    &n_r = 16\epsilon_*\eta_*-24\epsilon_*^2-2\xi_*\,,
    \qquad n_{rr} = -192 \epsilon_*^3 + 192\epsilon_*^2\eta_* - 32\epsilon_*\eta_*^2 - 24\epsilon_*\xi_* + 2\eta_*\xi_* + 2\sigma_*\,,
\end{aligned}
\ee
where the asterisk means that all the quantities are evaluated at the moment when the pivot scale, $k_* = 0.05\;\rm{Mpc^{-1}}$, exits the Hubble horizon.
These quantities are constrained by CMB observations. The most recent constraints on the scalar power spectrum are given by the Planck Collaboration \cite{Planck:2018vyg}: $\log \left(10^{10}A_s\right) = 3.047\,\pm\,0.014$, $n_s = 0.9647\,\pm\,0.0043$, $n_r = 0.0011\,\pm\,0.0099$, and $n_{rr} = 0.009\,\pm\,0.012$. The tensor-to-scalar ratio is constrained to be $r < 0.036$ by the BICEP/Keck Collaboration \cite{BICEP:2021xfz}.

The independent model parameters entering the calculation of the CMB power spectrum can be taken as $V_0$, $M$, $f$, and $a_*$.
Considering that $V_0 \gg M^4$, the observables are
\be
\begin{aligned}\label{eq:CMBObs}
  & r \simeq 8\,\left( \frac{\MP}{f}\frac{M^4}{V_0}\right)^2\sin^2\frac{a_*}{f}\,,\qquad && A_s \simeq\frac{1}{12\pi^2} \frac{f^2\,V_0^3}{\MP^6\, M^8}\sin^{-2}\frac{a_*}{f}\,,\qquad  n_s \simeq 1+2\,\frac{\MP^2}{f^2}\frac{M^4}{V_0}\cos \frac{a_*}{f}\,,\\ & n_r \simeq 2\,\left(\frac{\MP^2}{f^2}\frac{M^4}{V_0}\right)^2\sin^2\frac{a_*}{f}\,,\qquad &&n_{rr} \simeq -4\,\left(\frac{\MP}{f}\right)^6\left(\frac{M^4}{V_0}\right)^3\sin^2\frac{a_*}{f}\cos\frac{a_*}{f}\,,
\end{aligned}
\ee
while the number of $e$-folds of observable inflation is given by
\be\label{eq:efolds}
    N\simeq \frac{1}{\MP^2}\int_{a_{\rm end}}^{a_*}\frac{V_{\rm{eff}}}{V'_{\rm{eff}}} da\simeq \frac{f^2}{\MP^2} \frac{V_0}{M^4}\left(\log\tan\frac{a_*}{2f}-\log\tan\frac{a_{\rm end}}{2f}\right)\,.
\ee
We identify the field value $a_{\rm end}$ where inflation ends with the critical point $a_c = 6 f \alpha_c$, where the waterfall field becomes tachyonic; see Eq.~(\ref{eq:alphac}).
We will comment on the assumptions underlying this identification at the end of this section. As the end of inflation is controlled by the coupling between the inflaton and the waterfall field, $a_{\rm end}$ is an independent parameter of our model, which can be freely adjusted: By suitably choosing $a_{\rm end}$, we are always able to accommodate enough $e$-folds of inflation. For concreteness, in what follows, we fix $N=60$ as a reference value.

A characteristic feature of the model is to predict positive $n_r$ and $n_{rr}$. A second feature concerns the running of the running of the spectral index, which for this model is not independent from $n_s$ and $n_r$, $n_{rr} \simeq (1 - n_s)n_r$, leading to the constraint $n_{rr} < 4 \times 10^{-4}$. This is more stringent than the direct experimental bound on $n_{rr}$. The upcoming satellite mission SPHEREx \cite{SPHEREx:2014bgr} will measure the running of the spectral index with a target sensitivity $\Delta n_r \sim 10^{-3}$, and, therefore, it could rule out the model in the near future.

The four parameters of the model are not all determined by the CMB observables. Requiring that the model parameters do not suffer from fine-tuning problems will lead to further relations between them; see Sec.~\ref{sec:Naturalness}. To better understand how observation constrains them, we invert the relations given in Eq.~\eqref{eq:CMBObs} and obtain
\be
\begin{aligned}
    &V_0 = \frac{3\pi^2}{2}A_s r \MP^4\,,\qquad &&M^4 = \frac{3\pi^2}{16}\frac{A_s}{n_r}\sqrt{2 n_r+\left(n_s-1\right)^2}r^2 \MP^4\,,\\
    &f = \frac{1}{2}\sqrt{\frac{r}{n_r}} \MP\,,\qquad &&\frac{a_*}{f} = \arccos\left[\frac{n_s-1}{\sqrt{2n_r+(n_s-1)^2}}\right]\,.
\end{aligned}\label{eq:Par}
\ee

\begin{figure}[h!]
\begin{center}
\includegraphics[width=0.48 \textwidth]{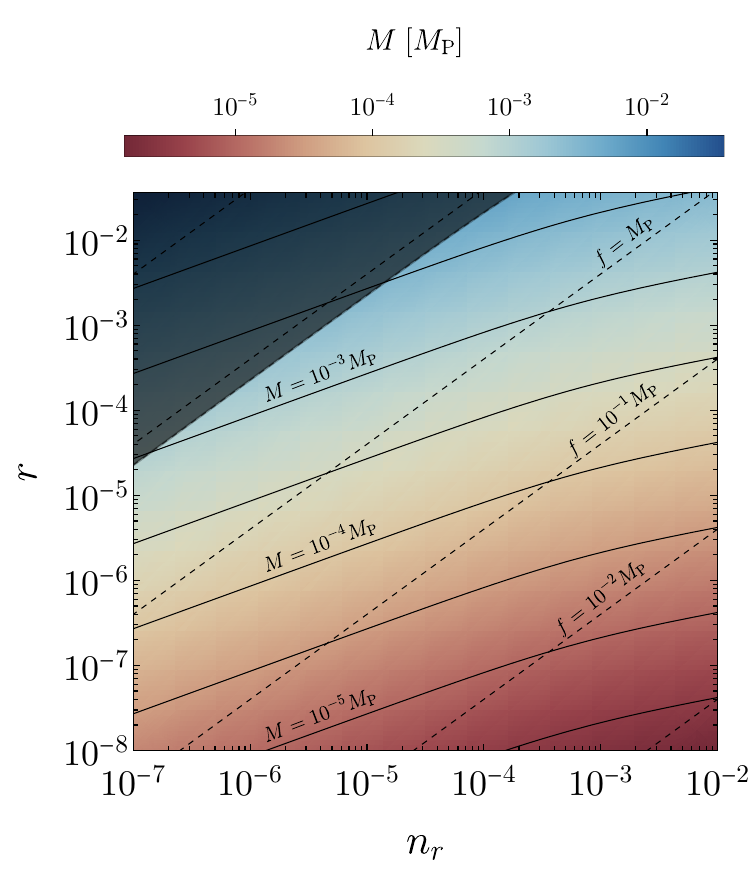}
\includegraphics[width=0.47\textwidth]{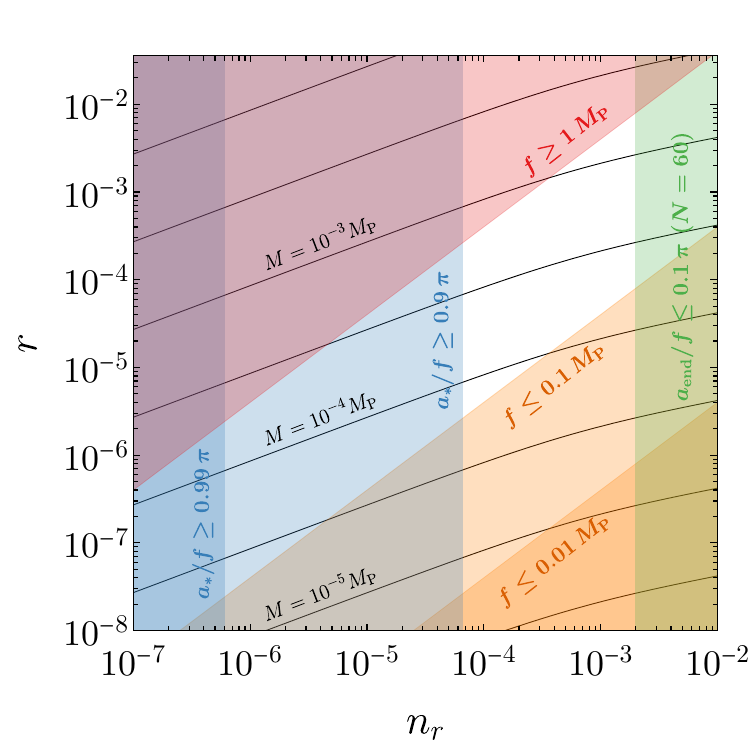}
\caption{
\textit{Left-hand panel.}
The values of $M$ and $f$ in units of $\MP$, as a function of $n_r$ and $r$,
while fixing $A_s = 2.105\times10^{-9}$ and $n_s = 0.965$. The largest displayed values of $n_r$ and $r$ correspond to the present experimental upper bounds. Solid (dashed) lines are contours of constant $M$ ($f$). The dark-gray region violates the slow-roll condition before the critical point.
\textit{Right-hand panel.} The white region is the one which complies with naturalness criteria, according to the discussion in Sec.~~\ref{sec:Naturalness}. In the orange regions the value of $f$ is unnaturally small, while in the red region $f>\MP$. The parameter space covered by the blue and green regions requires a significant fine-tuning of the inflation initial and final conditions, respectively.}
\label{fig:M} 
\end{center}
\end{figure}

By imposing $A_s=2.105\times10^{-9}$, $n_s=0.965$, $n_r < 0.011$, and $r<0.036$, we can derive the inequalities
\be\label{eq:Mup}
    V_0^{1/4}< 6\times 10^{-3}\MP\,,\qquad \frac{M^4}{f^4}\simeq  3\sqrt{2}\pi^2 A_s n_r^{3/2}< 10^{-10}\,.
\ee
In the left-hand panel in Fig.~\ref{fig:M}, we chart the region allowed by observations in the $(n_r,\,r)$ plane, once $A_s$ and $n_s$ are fixed. All the above relations have been derived under the assumption that $V_0$ drives inflation, i.e., $V_0 \gg M^4$. Below, we quantify what separation is needed between the two scales, in order not to end inflation before the critical point $a=a_c$. In particular, we require $\epsilon < 1$ all along the inflationary trajectory.\footnote{For slow-roll dynamics to persist, we also have to check that $\lvert\eta\rvert<1$ for any value of $a$, but this leads to a constraint which is very similar to Eq.~\eqref{eq:MV0ratio}.}
\be
\epsilon = \frac{\MP^2}{2}\left(\frac{\frac{M^4}{f} \sin{\frac{a}{f}}}{V_0 - M^4\cos{\frac{a}{f}}}\right)^2 < 1\,, \quad \forall\;a\,.
\ee
This is equivalent to requiring
\be
    \frac{M^4}{V_0}<\sqrt{\frac{2f^2}{2f^2+\MP^2}} \,,
\ee
or, equivalently,
\be\label{eq:MV0ratio}
    r<n_r\left(-1+\sqrt{1+\frac{64}{2n_r+(n_s-1)^2}}\right)\,.
\ee
The region where this condition is not satisfied is shaded in dark gray in the left-hand panel in Fig.~\ref{fig:M}.

\begin{figure}[h!]
\begin{center}
 \includegraphics[width=0.48\textwidth]{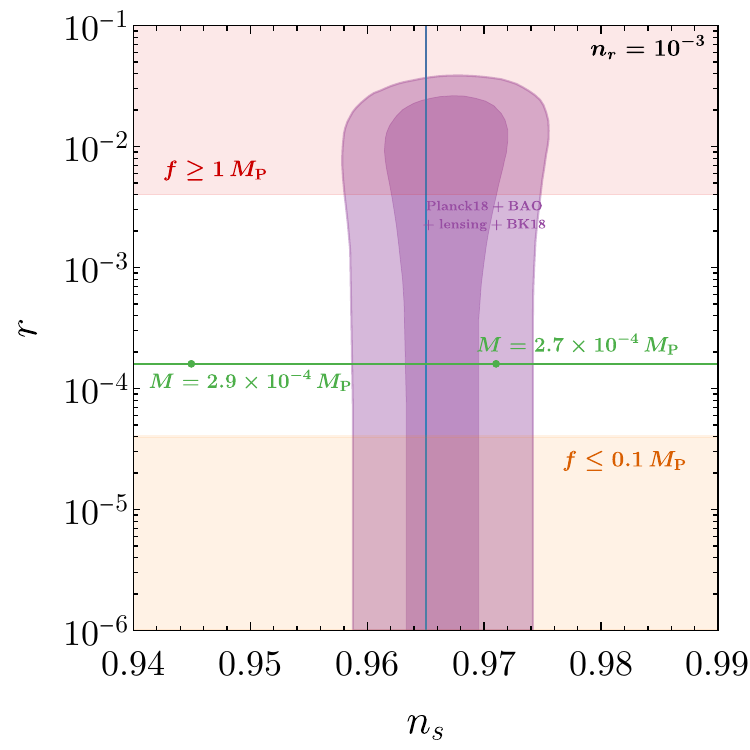} \includegraphics[width=0.48\textwidth]{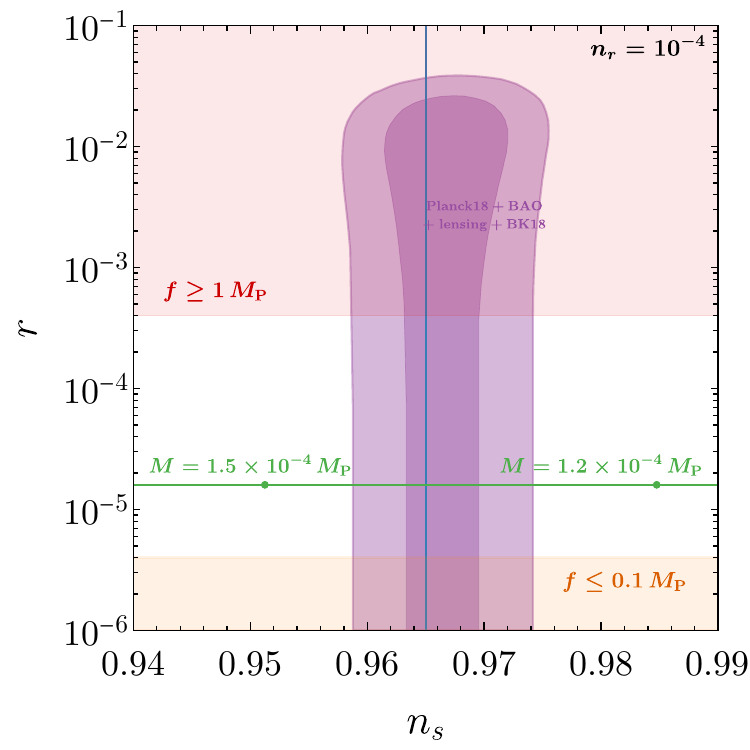}
\caption{The purple areas show the 95\% CL and 68 \% CL preferred regions from Planck 2018 + BAO + lensing + BICEP/Keck 2018 data \cite{BICEP:2021xfz}. We fixed the amplitude of the scalar power spectrum at $A_s = 2.105\times10^{-9}$. 
In the \textit{left-hand panel} we set the running of the spectral index at $n_r = 10^{-3}$, while in the \textit{right-hand panel} we took $n_r=10^{-4}$.
The horizontal, green lines are obtained by fixing $f=0.2\,\MP$ and varying $M$, while along the vertical, blue lines $M/f=3\times 10^{-3}$ is fixed and $f$ varies. The orange and the red regions are disfavoured by naturalness (see section~\ref{sec:Naturalness}): Low values of the tensor-to-scalar ratio $r$ require unnaturally small $f$, while a large $r$ is associated to transplanckian $f$. The relation between $f$ and $r$ depends on the value assumed for $n_r$, according to Eq.~\eqref{eq:Par}.}
\label{fig:Planck} 
\end{center}
\end{figure}

In Fig.~\ref{fig:Planck}, we show the predictions of the model in the $(n_s,\,r)$ plane, together with the region allowed by the latest CMB measurements \cite{Planck:2018jri, BICEP:2021xfz}. Our model is in agreement with observations for natural values of $f$, i.e., for $f$ smaller than, but not too far from $\MP$.

We require the end of inflation to occur within one Hubble time after the critical point has been reached by the accident. This allows us to identify $a_c = 6f\alpha_c$ with the point $a_{\rm end}$ in field space in which inflation ends. For this requirement to hold, the Hubble rate at the critical point, $H_c$, must be much lower than the value of the waterfall mass, after $\Delta t \sim H_c^{-1}$ has elapsed since $a = a_c$:
\be
    \lvert m_{\chi^3}^2\left(a_c + \Delta a\right)\rvert \gg H_c^2\,,
\ee
where $\Delta a$ is the accident excursion during $\Delta t$. 
This condition is also known as the ``waterfall condition'' \cite{Linde:1993cn}. Given that the energy density during inflation is dominated by the false vacuum, the Hubble rate can be approximated to $H_c \simeq V_0/3 \MP^2$. Moreover, the evolution of $a$ is governed, in the slow-roll approximation, by $3\,  H_c\,\dot{a} \simeq \partial V_{\rm eff}/\partial a$. With these considerations in mind, the above waterfall condition 
translates into a lower bound on the inflaton-waterfall quartic coupling:
\be \label{eq:WaterCond}
    \zeta \gg  \frac{V_0^2}{18\,M^4 \MP^4}\,\frac{1}{\sin\left( a_c/f\right) \sin\left( a_c/3f \right)}\,.
\ee

By making use of the relations in Eqs.~\eqref{eq:Par} and \eqref{eq:efolds}, we find that the right-hand side of the above condition is a monotonic, increasing function of $n_r$, once $A_s$, $n_s$, and $N$ are fixed. For $n_r$ taking its largest value allowed by observations, Eq.~\eqref{eq:WaterCond} becomes $\zeta \gg 10^{-6}$, meaning that we can always choose a perturbative value of $\zeta$ for which the waterfall condition holds.
We note that the requirement for inflation to end as soon the critical point is reached is not imposed by observations. In principle, one could also allow inflation to proceed along the waterfall direction for some 20 $-$ 40 $e$-folds as realized in the so called ``mild hybrid inflation'' scenario \cite{Clesse:2010iz, Clesse:2015wea,Kawasaki:2015ppx}.

\subsection{Naturalness considerations}\label{sec:Naturalness}

The aim of this section is to assess which portion of the parameter space allowed by observations complies with naturalness. An important motivation for considering this model in the first place is that the inflaton potential is nearly flat for generic values of the couplings, without requiring fine-tuning. To what extent does the model remain ``natural'', or minimally fine-tuned, once we impose that it should give rise to a quantitatively realistic phenomenology?

It turns out that there is a preferred range of natural values for the running of the spectral index $n_r$. A lower bound on $n_r$ is obtained as follows: From Eq.~\eqref{eq:CMBObs}, we have
\be
    n_r = \frac{1}{2}\left(n_s-1\right)^2 \tan^2 \frac{a_*}{f}\,.
\ee
Hence, fixing $n_s$ at its value preferred by CMB measurements, $n_r$ could be arbitrarily small, but only at the cost of tuning the start of observable inflation $a_*$ arbitrarily close to $a_*=\pi f$. However, when remaining agnostic about any pre-inflationary physics, the inflaton could start rolling anywhere in field space. Taking, for instance, 10\% as an acceptable degree of fine-tuning of the initial condition,
\be
    \frac{a_*}{\pi f} \lesssim 0.9\,,
\ee
one obtains the naturalness bound $n_r\gtrsim 6.5\times10^{-5}$. Requiring the fine-tuning to be less than 1\% gives, instead, $n_r \gtrsim 6\times 10 ^{-7}$. Such lower bounds are shown as blue-shaded regions in the right-hand panel in Fig.~\ref{fig:M}.

An upper bound on $n_r$ is similarly given by the following argument: Fixing the number $N$ of observable e-folds to $N=60$, the parameter combination $f^2 V_0/(\MP^2 M^4)$ could, in principle, be made arbitrarily small by tuning the ending point of inflation $a_{\rm end}$ close to zero; see Eq.~\eqref{eq:efolds} (and by simultaneously tuning $a_*$ close to $a_*=\frac{\pi}{2} f$ to obtain the observed value of $n_s$ according to Eqs.~\eqref{eq:CMBObs}). This would imply, again by Eqs.~\eqref{eq:CMBObs}, that $n_r$ could be arbitrarily large. But $a_{\rm end}$ is given by a combination of independent model parameters which have no reason to conspire to produce $a_{\rm end}\simeq 0$. Imposing for concreteness
\be
\frac{a_{\rm end}}{\pi f} \gtrsim 0.1\,,
\ee
this yields $n_r\lesssim 2\times10^{-3}$, shown by green shading in the right-hand panel in Fig.~\ref{fig:M}. 

Moreover, naturalness requires $f$ to be not too far from the cutoff scale $\Lambda_{\rm UV}$ of the model. This is nothing but the well-known hierarchy problem of models with elementary scalar fields: Quadratically divergent loop corrections to the mass parameters $\mu_\phi^2$ and $\mu_\chi^2$ signal that they depend sensitively on the ultraviolet dynamics at the scale $\Lambda_{\rm UV}$. At this scale, the model might be embedded in a supersymmetric field theory, or in a strongly coupled field theory where $\phi$ and $\chi$ emerge as bound states. However, the most minimal assumption is that there is no field-theoretic embedding which cures the UV sensitivity, in which case we should identify $\Lambda_{\rm UV}$ with the quantum gravity scale.

Quantum gravity corrections are, of course, greatly model-dependent. Identifying $\Lambda_{\rm UV}=\MP$ for definiteness, and assuming that UV states have $\lesssim {\cal O}(1)$ couplings to $\phi$ and $\chi$, we can derive a tentative naturalness bound on the tensor-to-scalar ratio $r$. Cutting off the one-loop corrections to the $\mu_\phi^2$ parameter at the scale $\MP$ and requiring that they do not exceed the tree-level value leads to
\be \mu_\phi^2 \gtrsim\left(7\,\lambda_\phi + \frac{3}{2}\,\varepsilon + 3\,\zeta \right)\frac{\MP^2}{16\pi^2}\,.
\ee
Since $f^2=\mu_\phi^2/6\lambda_\phi$, this implies $f\gtrsim 0.1\,\MP$, under the technically natural assumption that $\varepsilon$ and $\zeta$ are at most of the order of $\lambda_\phi$. 
Therefore, the orange-shaded region in the right-hand panel in Fig.~\ref{fig:M} is disfavoured.
From Eqs.~\eqref{eq:CMBObs}, it follows that $r=4\,n_r\,(f/\MP)^2$ and, thus,
\be
    r \gtrsim (4\times 10^{-2})\,n_r\,,
\label{eq:NatBound}
\ee
shown in orange shading in Fig.~\ref{fig:Planck}. Depending on how much fine-tuning one is willing to accept, this bound can, of course, shift by several orders of magnitude, as also indicated in the right-hand panel in Fig.~\ref{fig:M}.

On the other hand, values of $f$ too close to $\Lambda_{\rm UV}$ imply that higher-dimensional operators, which may spoil the tree-level flatness of the inflaton potential, can no longer be neglected. In that sense, the ``least fine-tuned'' region of parameter space is also the one with the least theoretical control. In Figs.~\ref{fig:M} and \ref{fig:Planck}, we shaded in red the regions with $f\gtrsim \MP$, or, equivalently, $r\gtrsim 4n_r$.

Such naturalness constraints translate into some bounds on the model parameters. 
For instance, the inflationary scale is given by $V_0 \simeq \lambda_\chi v_\chi^4/4$, 
and the above bounds on $r$ and $n_r$ can be translated into bounds on $V_0$ once $A_s$ is fixed, according to Eq.~\eqref{eq:Par}. This, together with $0.1\,\MP \lesssim v_\chi \lesssim \MP$ (we require $v_\chi$ to be subplanckian, analogously to $f$), gives a natural range of values for the $\chi$ quartic coupling:
\be
   10^{-13} \lesssim \lambda_\chi \lesssim 10^{-5}
    \,.
\ee

\section{Gravitational waves from preheating}

Hybrid inflation is followed by a stage of tachyonic preheating \cite{Felder:2000hj, Felder:2001kt}. The tachyonic instability of the waterfall field exponentially enhances field inhomogeneities. Such inhomogeneous, bubble-like structures behave as scalar waves and scatter off each other until the Universe is homogenized and driven to local thermal equilibrium.  A stochastic gravitational wave background (SGWB) is produced in the process  \cite{Garcia-Bellido:2007fiu,Garcia-Bellido:2007nns, Dufaux:2007pt, Dufaux:2008dn}. GWs quickly decouple from the background dynamics and propagate (almost) undisturbed throughout cosmic history, hence carrying information about the process that sourced them. Given the energy scale of inflation, which is typically large, the frequency at which GWs are produced during preheating lies well beyond the reach of current or even planned interferometers. However, it has been shown, e.g., in Ref.~\cite{Dufaux:2008dn}, that, by sufficiently lowering the scale of inflation, the signal might be brought within the projected sensitivity of the Einstein Telescope (ET) \cite{Maggiore:2019uih}. It is also worth mentioning that there have been proposals of using optical atomic clocks to detect high-frequency GWs; see, e.g., Ref.~\cite{Bringmann:2023gba} and references therein.

The spectrum of GWs produced during preheating presents a peaked shape, with the frequency and amplitude at the peak given, respectively, by \cite{Dufaux:2008dn}
\be\label{eq:TachPre}
    f_\star \sim 4\times 10^{10}\,{\rm Hz} \frac{k_\star}{{\rho_c}^{1/4}}\,,\qquad h^2\Omega_\star \sim 10^{-6} \left(\frac{H_c}{k_\star}\right)^2\,,
\ee
where $k_\star$ is the typical momentum enhanced by the tachyonic instability, related to the characteristic size of inhomogeneities as $R_\star \sim 1/k_\star$.\footnote{Here $k_\star$ is unrelated to the pivot scale of CMB observables $k_*$ in Sec.~\ref{sec:CMB}.} Here, $\rho_{c}\simeq V_0$ and $H_c^2\simeq V_0/(3\MP^2)$ are, respectively, the energy density and the Hubble rate at the end of inflation or, analogously, at the onset of the tachyonic instability. 
The value of $k_\star$ depends on the inflationary dynamics close to the waterfall transition.

From Eq.~\eqref{eq:TachPre}, one may already deduce that, for a signal of GWs from preheating to be within the reach of the ET, the inflation scale must lie in the range
\be
 3\times 10^5\,{\rm GeV}\lesssim V_0^{1/4}\lesssim 5\times 10^{11}\,{\rm GeV}\,,
\ee
which is associated to an unobservably small tensor-to-scalar ratio, according to Eq.~\eqref{eq:Par}. Should $r$ be measured in the future, along with a SGWB at the ET which can be traced back to tachyonic preheating, this would, therefore, rule out our model (and other models of hybrid inflation, where tachyonic preheating is an unavoidable feature).

In principle, two different regimes for GW production from preheating can be identified  \cite{Dufaux:2008dn}, depending on the velocity $\dot{a}_c$ of the inflaton at the critical point $a=a_c$ where the waterfall field becomes tachyonic. In the first regime, $a$ is fast enough for the tachyonic instability to be driven by its classical dynamics. Another possibility is that $\dot a_c$ is sufficiently small, so that preheating is triggered by quantum fluctuations. In the present work, we take model parameters such that inflation ends as soon as the inflaton crosses the critical point; see Eq.~\eqref{eq:WaterCond}. In that case, classical rolling always dominates.

\begin{figure}[h!]
\begin{center}
\includegraphics[width=0.48\textwidth]{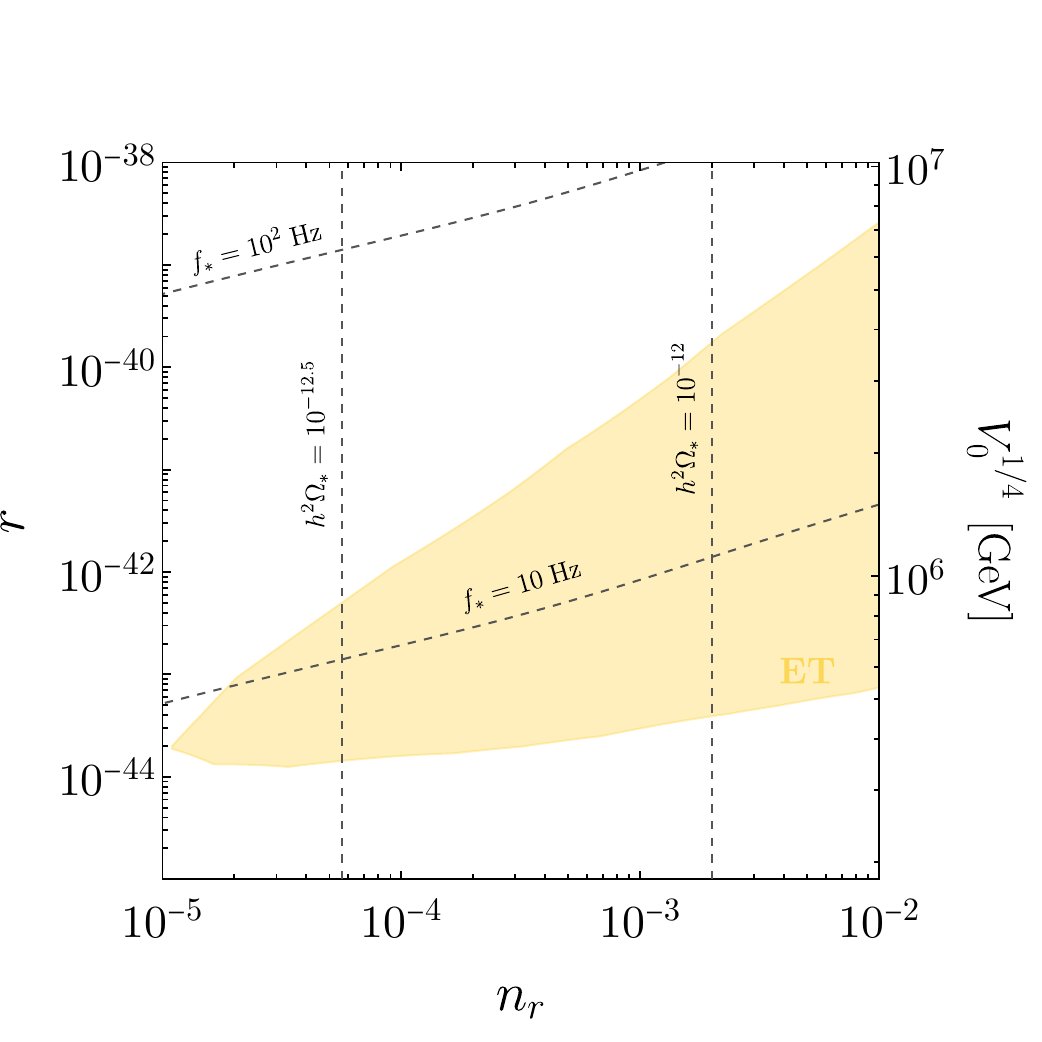}
 \includegraphics[width=0.48\textwidth]{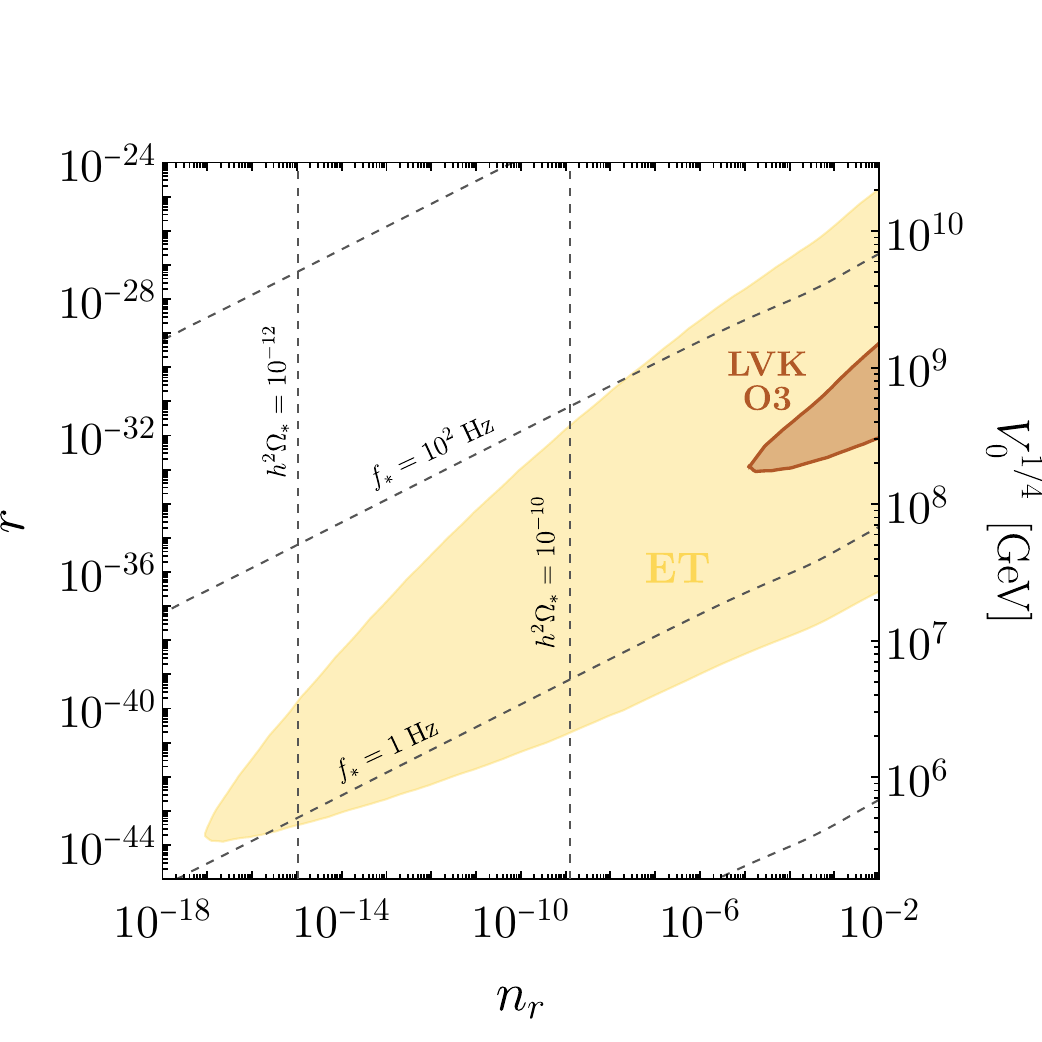}
    \caption{Sensitivity region (yellow) of the ET \cite{Maggiore:2019uih} for the GW spectrum produced during tachyonic preheating, as a function of $n_r$ and $r$ (or equivalently $V_0$).
    In the \textit{left-hand panel} we take $\tilde{\mu}_\chi^2/f^2 = 1$, while in the \textit{right-hand panel} we fix $\tilde{\mu}_\chi^2/f^2 = 10^{-6}$. 
    The projected sensitivity for the ET is the one derived in Ref.~\cite{NANOGrav:2023hvm}, and it is shown in Fig.~\ref{fig:DWGW}.
    To guide the eye, we also plot lines of constant peak frequency and amplitude for the preheating GW spectrum.
    The brown region shows the integrated sensitivity for the third observing run of the LIGO-Virgo-KAGRA network \cite{KAGRA:2021kbb}.}
\label{fig:ET}  
\end{center}
\end{figure}

We are interested in an order-of-magnitude estimate of the amplitude and frequency of the GWs signal at the peak. Because of the fact that tachyonic preheating is a non-linear process, a more precise computation of the GW spectrum would require lattice simulations which are beyond the scope of this work. In the slow-roll approximation, $\ddot{a} \ll 3H\dot{a}$, the inflaton velocity at the critical point reads
\be
    \dot{a}_c \simeq \frac{\MP}{\sqrt{3}f}\frac{M^4}{V_0^{1/2}}
    \sin\frac{a_c}{f}\,.
\ee
If its velocity is large enough, the inflaton will keep rolling classically past the critical point. At $a=a_c$, the mass-squared of the waterfall field vanishes and then starts growing negative so that after an interval $\Delta t$ it becomes $m_{\chi^3}^2 \simeq 6 \, \zeta \, f \, \dot{a}_c\, \Delta t \,\sin \left(a_c/3f\right)$. The exponential growth of quantum fluctuations becomes efficient when the tachyonic mass of the waterfall field has grown large enough such that $\sqrt{|m_{\chi^3}^2|} \gtrsim \Delta t^{-1}$. This leads to the enhancement of the mode $k_\star \sim 1/\Delta t \lesssim \sqrt{|m_{\chi^3}^2|}$, so that
\be\label{eq:kCl}
    k_{\star,{\rm cl}}^3 \simeq 2\sqrt{3}\,\zeta\,\frac{M^4 \MP}{V_0
^{1/2}}\sin\frac{a_c}{f}\sin\frac{a_c}{3f}\,.
\ee
Fig.~\ref{fig:ET} shows the values of $r$ (or, analogously, of the inflationary scale $V_0$) and $n_r$ which correspond to a GW signal within the ET projected sensitivity for $\tilde{\mu}_{\chi}^2/f^2 = 1$ (left-hand panel) and $\tilde{\mu}_{\chi}^2/f^2 = 10^{-6}$ (right-hand panel), where the ratio $\tilde{\mu}_{\chi}^2/f^2$ measures the steepness of the waterfall direction.
 
We conclude this section by mentioning that, in principle, tachyonic instability could also lead to production of primordial black holes (PBHs).\footnote{As a side note, we touch on the possibility of producing PBHs at the end of inflation thanks to an enhancement of curvature perturbations driven by a large running of the spectral index. It has been shown in Ref.~\cite{Kohri:2007qn} that this eventuality bounds $n_r \lesssim 10^{-2}$ in single-field, slow-roll inflation. However, larger values for $n_r$ are nowadays excluded by CMB observations.} In our model, however, inflation ends almost immediately after reaching the critical point. This leads to enhancement of perturbations at very small scales and to a subsequent production of extremely light PBHs which have completely evaporated by now. It is possible to envision a scenario in which inflation goes on for a certain number of e-folds along the waterfall direction, hence producing PBHs which can be of phenomenological relevance \cite{Afzal:2024xci, Braglia:2022phb}.

\section{Topological defects}\label{sec:TopologicalDefects}

Models with spontaneously broken symmetries may give rise to topological defects such as magnetic monopoles, cosmic strings (CSs), and domain walls (DWs). 

Magnetic monopoles can appear if the second homotopy group of the  vacuum manifold is nonzero. When $G=\SO{3}$ is broken to $\SO{2}$ by a triplet VEV, as is the case in our model, this gives rise to the celebrated 't Hooft-Polyakov monopole \cite{tHooft:1974kcl, Polyakov:1974ek} with $\pi_2(\SO{3}/\SO{2})= \pi_2[S^2]=\mathbb{Z}$.  However, in our models, the relevant phase transition takes place before inflation, and, therefore, no monopoles are likely to remain in our Hubble patch. To be precise, symmetry breaking in our model proceeds in two steps: First, at the scale $v$, the symmetry group $G$ is broken completely by the 5-plet VEV, while the triplet VEV is zero in this phase. The vacuum manifold (or rather the scalar field manifold which minimizes the potential for all fields except the slowly-rolling inflaton) is $G$, and in different causally connected regions of the Universe, the scalar fields will take VEVs in different domains of $G$. However, the subsequent phase of inflation stretches out these inhomogeneities, such that, at all scales relevant to us, the $5$-plet VEV becomes effectively a single point in field space. Nontrivial field configurations which might evolve into monopoles are ``inflated away''. The second step of symmetry breaking happens at the waterfall transition at the end of inflation, when the $5$-plet VEV relaxes to zero and a triplet VEV is generated, thereby restoring an $\SO{2}$ symmetry. The vacuum manifold becomes $\SO{3}/\SO{2}\simeq S^2$; however, no new monopoles can be created, since this would require energies sufficiently high to restore the entire $\SO{3}$ symmetry. We conclude that, for better or worse, magnetic monopoles will not affect the cosmology of our model.

Cosmic strings can appear if the scalar vacuum manifold is not simply connected. This is not the case in the minimal model, but it is interesting to consider a non-minimal extension to study the phenomenological consequences. Consider a version of the model in which the triplet $\chi$ is a complex scalar field transforming under $G=\SU{2}\times\U{1}_\chi$, with $\U{1}_\chi$ acting as a phase rotation. All or part of $G$ may be thought of as gauged. We can further extend the model by promoting also $\phi$ to a complex field and gauging $\U{1}_\phi$, which is appealing because then the model no longer relies on any ad-hoc discrete symmetries, the role of the $\mathbb{Z}_2$ in the original model now being played by $\U{1}_\phi$. The technical details of these model versions are worked out in Appendix \ref{sec:AppendixComplexModel}.  These models do allow for additional quartic scalar couplings, but the qualitative features of the inflaton potential, as well as CMB predictions, do not change: Crucially, there is still an accidentally light scalar field which can play the role of the inflaton, with a cosinusoidal loop-induced potential analogous to Eq.~\eqref{eq:CosPot}. Likewise, the effective mass parameter of the waterfall field $\chi$ depends on the value of the inflaton, and parameters can be chosen such that it becomes tachyonic at the end of inflation. The waterfall field is zero along the inflationary trajectory, so $\U{1}_\chi$ remains unbroken during inflation while $\SU{2}\times\U{1}_\phi$ is completely broken. The true vacuum is at $\vev{\phi}=0$ and $\vev{|\chi|}=\vev{|\chi^3|}=v_\chi\neq 0$, so the vacuum manifold is $S^2\times\U{1}_\chi$. 

As $\pi_1[U(1)]=\mathbb{Z}$, the SSB phase transition results in the production of a network of stable, local CSs with a typical tension \cite{Vilenkin:2000jqa}
\be
    \GN\mu \simeq 2\pi\, \GN\, v_{\chi}^2\,,
\ee 
where $v_{\chi}$ is the breaking scale of $\U{1}_{\chi}$. It is well known that this network quickly reaches a scaling regime, where the fraction of energy density stored in the CSs redshifts in the same way as the energy density of the background \cite{Hindmarsh:1994re,Vilenkin:2000jqa}. Strings can intercommute and produce loops, which oscillate under their own tension and radiate energy in the form of GWs. Such a signal of GWs appears to us as a SGWB with a characteristic scale-invariant spectrum (if standard cosmology is assumed \cite{Gouttenoire:2019kij}). The peak and the amplitude of the GWs signal for stable strings are both given by the string tension, which, in order to match the signal recently reported by PTA collaborations, has to be $\GN\, \mu \simeq 10^{-10}$ \cite{NANOGrav:2023hvm,EPTA:2023xxk}, even though stable CSs seem to be disfavoured with respect to other cosmological sources \cite{NANOGrav:2023hvm}. The reason is that, for stable CSs, both the amplitude and the shape of the GW signal are controlled by $\GN\mu$; once we fix the string tension to match the amplitude of the reported signal, the CS spectrum turns out to be flat in the nHz region, with data preferring a blue-tilted spectrum. $\GN\mu \simeq 10^{-10}$ represents a trade-off between a signal which is not too weak and one which is not too flat.  
However, as the computation of the expected signal of GWs from a CS network is subject to large uncertainties, we can take as upper bound on the string tension the one given by the CMB constraint on the impact of CSs on the CMB power spectrum \cite{Planck:2015fie}, $\GN\mu \lesssim 10^{-7}$, which corresponds to an inflationary scale $V_0^{1/4}\lesssim \left(\lambda_\chi + \lambda_\chi'\right) \left(3\times 10^{14}\right)\,{\rm GeV}$.

Finally, domain walls can appear if the scalar vacuum manifold is not connected. This, again, is not the case in our minimal model. In the following, we exploit a non-minimal model with disconnected minima,
which is described in detail in Appendix \ref{sec:AppendixZ4Model}.
In this case, the PTA signal can be fit with GWs produced from annihilating DWs, as explained in the next section. The waterfall field $\chi$ is taken to be a complex scalar triplet,
subject to a discrete symmetry $\mathbb{Z}_{4\chi}: \chi\into i\chi$.
This leads to disconnected minima which are degenerate, 
corresponding to stable DWs.
However, the vacuum degeneracy can be easily lifted by
softly breaking the $Z_{4\chi}$ discrete symmetry, allowing for the DWs to annihilate. Also in this model the inflationary predictions are the same as those for the minimal model.

\subsection{Gravitational waves from unstable domain walls}\label{secDW}

In the model described in Appendix \ref{sec:AppendixZ4Model}, after the end of inflation the $\chi$ scalar potential has two disconnected, degenerate minima, so that DWs are formed with a surface energy density \cite{Vilenkin:2000jqa}
\be\label{eq:DWTens}
    \sigma = \frac{2\sqrt{2}}{3}\lambda_{\chi}^{1/2}v_{\chi}^3\,,
\ee
where $\lambda_\chi$ is the $\chi$ quartic coupling and $v_\chi$ its VEV.
The motion of DWs is initially damped by their interaction with the surrounding plasma. Numerical simulations show that, after some time, friction becomes negligible and the DW network reaches the ``scaling regime'' \cite{Saikawa:2017hiv} in which their energy density scales with time as $\rho_{\rm DW} = \sigma \mathcal{A}/t$, with $\mathcal{A} \simeq 0.8$ \cite{Hiramatsu:2013qaa}.
If the vacuum degeneracy were exact, the DWs would be stable and they would quickly overclose the Universe, as,  in the scaling regime, $\rho_{\rm DW}$ redshifts more slowly than the energy density of the background. DWs start dominating when $3H^2\MP^2 = \rho_{\rm DW}$, which for a previously radiation-dominated universe corresponds to the time
\be
    t_{\rm dom}^{\rm DW} = \frac{3\, \MP^2}{4\,\mathcal{A}\,\sigma}\,.
\ee
However, the discrete symmetry ensuring the vacuum degeneracy may be softly broken, which lifts the degeneracy and introduces a potential bias $\Delta V$; see Eq.~\eqref{eq:Vbias}. Provided that $\Delta V<0.795\,V_0$ \cite{Saikawa:2017hiv}, where $V_0$ is the potential energy at $\chi=0$, DWs are formed, but they are unstable as long as $\Delta V\neq 0$. The different domains annihilate when the bias pressure $\Delta V$ dominates over the DW surface tension at
\be\label{eq:tann}
    t_{\rm ann} \simeq C_d \mathcal{A} \frac{\sigma}{\Delta V}\,,
\ee
with $C_d \simeq 3$ \cite{Gouttenoire:2023ftk}, until all the observable Universe lies in the true vacuum. We require the DW network to disappear before the bias energy density of the false-vacuum patches dominates the Universe:
\be
    t_{\rm dom}^{\Delta V} \simeq \frac{\sqrt{3}\MP}{2\sqrt{\Delta V}}\,.
\ee
The requirement $t_{\rm ann} < t_{\rm dom}^{\Delta V}$ leads to the constraint $\Delta V \gtrsim 8\,\sigma^2/\MP^2$. 

In an early phase, as far as we know, the Universe might well have been dominated by DWs, provided that they annihilate before big bang nucleosynthesis (BBN). However, the dynamics of a DW-dominated Universe is not well understood, and its treatment requires dedicated numerical studies \cite{Saikawa:2017hiv}, so we restrict ourselves to a radiation-dominated scenario. By combining this requirement together with that of having a not too large $\Delta V$, in order to have DWs appear in the first place, we obtain $v_\chi \lesssim 0.3\,\MP$, where we have made use of Eq.~\eqref{eq:DWTens} and $V_0 = \lambda_\chi v_\chi^4$.

As the DW network evolves, the DWs are driven to relativistic speed and radiate GWs. The signal of GWs, appearing to us as a SGWB, has a peak frequency given by the Hubble parameter at the moment of annihilation, which after taking into account redshift becomes \cite{Saikawa:2017hiv}
\be\label{eq:fpDW}
    f_{p}\simeq \frac{R\left(t_{\rm ann}\right)}{R\left(t_0\right)}
    H_{\rm ann}\simeq  1.6 \times 10^{-7}\;{\rm Hz} \left(\frac{g_*(T_{\rm ann})}{100}\right)^{1/6}\frac{T_{\rm ann}}{ {\rm GeV}}\,.
\ee
Here $R\left(t_{\rm ann}\right)$ and $R\left(t_0\right)$ are the scale factors at the time of DW annihilation and today, respectively,
$g_*$ is the number of effective degrees of freedom in relativistic energy, and $T_{\rm ann}$ is the temperature of the Universe at the moment of DW annihilation:
\be\label{eq:Tann}
    T_{\rm ann} = \left[\frac{45}{2\pi^2g_*(T_{\rm ann})}\;\frac{\MP^2}{t_{\rm ann}^2}\right]^{1/4} \,.
\ee
The present energy density of radiated GWs is \cite{Saikawa:2017hiv}
\be\label{eq:OmegaDW}
    h^2\Omega_{\rm GW}\left(f\right) \simeq 1.6\times10^{-5}\left(\frac{100}{g_*(T_{\rm ann})}\right)^{1/3}\, \frac{3}{32 \pi}\, \tilde{\epsilon}\, \alpha_{\rm ann}^2\,\mathcal{S}(f/f_p),
\ee
where $\tilde{\epsilon} =0.7$ is an efficiency parameter taken from simulations \cite{Hiramatsu:2013qaa},  $\alpha_{\rm ann}$ is the DW network energy density fraction at the time of annihilation, 
\be\label{eq:AlphaAnn}
    \alpha_{\rm ann} \equiv \frac{\rho_{\rm DW}(t_{\rm ann})}{\rho_r(t_{\rm ann})} \simeq \frac{4}{3}\,C_d\,\mathcal{A}^2\frac{\sigma^2}{\MP^2\,\Delta V}\,,
\ee
and the spectral function $\mathcal{S}(x)$ can be modeled as \cite{NANOGrav:2023hvm}
\be
    \mathcal{S}(x) = \left(\frac{a+b}{b\,x^{-a/c}+a\, x^{b/c}}\right)^c\,.
\ee
Numerical simulations have shown that $b,c\simeq 1$ for $\mathbb{Z}_2$-type DWs \cite{Hiramatsu:2013qaa}, and our model falls in this category, while causality fixes $a=3$ \cite{Hook:2020phx}.\footnote{In principle, the causality tail of the GW spectrum is affected by the QCD phase transition and the subsequent change in the number of relativistic degrees of freedom. This leads to a deviation from the pure power-law behaviour with $a=3$. Including the effect of the QCD phase transition is beyond the scope of this work, and we refer to Ref.~\cite{Franciolini:2023wjm} for a detailed discussion on the topic.}  By plugging Eqs.~\eqref{eq:tann} and \eqref{eq:Tann} into Eq.~\eqref{eq:fpDW}, we observe that the GW peak frequency scales with the DW network parameters as $f_p\sim\sqrt{\Delta V/\sigma}$. Similarly, by plugging Eq.~\eqref{eq:AlphaAnn} into Eq.~\eqref{eq:OmegaDW}, we find that $h^2 \Omega_{\rm GW}(f_p) \sim (\sigma^2/\Delta V)^2$.

\begin{figure*}[tbp]
\begin{center}
      \includegraphics[width=16 cm]{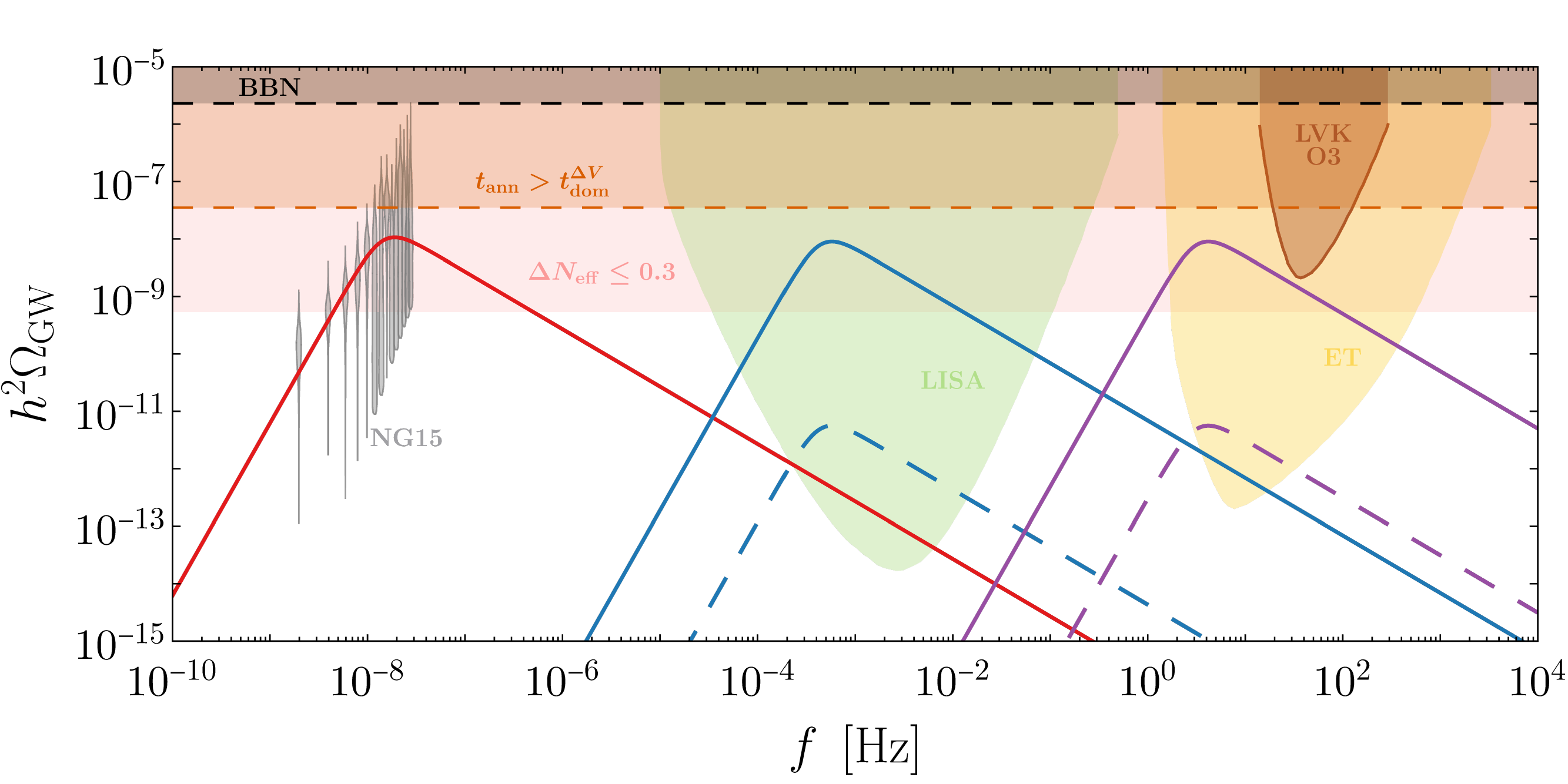}
    \caption{GW spectra produced by unstable DWs for different benchmark points. The red curve is obtained with DW tension $\sigma^{1/3} =3.2\times10^{5}\,{\rm GeV}$ and bias $\Delta V^{1/4} =0.23\,{\rm GeV}$. The blue solid curve corresponds to $\sigma^{1/3} = 3.2\times10^8\,{\rm GeV}$, and the purple solid one to $\sigma^{1/3} =1.2\times10^{11}\,{\rm GeV}$ ($\Delta V$ is varied accordingly to keep the amplitude of the signal fixed). The dashed spectra are obtained from the solid ones by decreasing $\sigma$ while keeping $f_p$ fixed: The dashed blue curve corresponds to $\sigma^{1/3} = 9.3\times 10^7\,{\rm GeV}$, while the dashed purple one to $\sigma^{1/3} = 3.5\times 10^{10}\,{\rm GeV}$. The gray violins show the SGWB reported by the NANOGrav 15 yrs experiment \cite{NANOGrav:2023hvm}. The green and yellow shaded regions show the projected sensitivities of LISA \cite{LISACosmologyWorkingGroup:2022jok} and ET \cite{Maggiore:2019uih}, respectively, as derived in Ref.~\cite{NANOGrav:2023hvm}, assuming a signal-to-noise-ratio equal to unity and an observing time of one year. The integrated sensitivity of the  LIGO-Virgo-KAGRA network during the third observing run \cite{KAGRA:2021kbb} is shown in brown. The region where the bias energy density dominates the Universe, leading to local departures from radiation domination, is shown in orange, while the black dashed line shows the bound on the amplitude of primordial GWs given by BBN \cite{LISACosmologyWorkingGroup:2022jok}. If the DW network decays into dark radiation, the pink region is excluded by the BBN bound on the effective number of neutrino species.
    }
\label{fig:DWGW} 
\end{center}
\end{figure*}

By taking the quartic couplings to be of the order of one, the DW tension is related to the inflationary scale as $V_0 \simeq \sigma^{4/3}$. This allows us to relate the GW signal generated by the DW network to inflation. It has been shown \cite{NANOGrav:2023hvm,Gouttenoire:2023ftk,Ellis:2023oxs} that a network of unstable DWs is a good candidate to explain the signal of a SGWB recently detected by the NANOGrav Collaboration \cite{NANOGrav:2023hvm}. We find that a model with $V_0^{1/4} \simeq 2\times10^{5}\, {\rm GeV}$ and a suitable bias $\Delta V$ leads to a GW signal which fits the NANOGrav detection. The peak of the GW signal is within the reach of the upcoming space-borne Laser Interferometer Space Antenna (LISA) \cite{LISACosmologyWorkingGroup:2022jok} for $3\times10^7\,{\rm GeV}\lesssim V_0^{1/4}\lesssim 2\times10^{10}\,{\rm GeV}$, while for $3\times10^{10}\,{\rm GeV} \lesssim V_0^{1/4}\lesssim 7\times 10^{12}\,{\rm GeV}$ the SGWB might be detected by the ET in the future. Fig.~\ref{fig:DWGW} shows several different benchmark GW spectra together with the sensitivity curves of current and planned experiments.

The energy density released by DW annihilation can potentially alter the expansion rate of the Universe. In order not to spoil BBN, the products of DW annihilation have to decay before the BBN epoque, either into dark radiation (DR) or into Standard Model (SM) particles  \cite{Ferreira:2022zzo}. Indeed, $\chi$ can decay into the $\SO{2}$ dark photon, or into SM degrees of freedom via, e.g., a Higgs-portal coupling. The dominant decay channel depends on the relative magnitude of the couplings and is related to the reheating mechanism, which we do not explore in this paper. 
The usual assumption, which also leads to the weakest constraints, is that such decays are already active at $T_{\rm ann}$ \cite{Ferreira:2022zzo}. Assuming the DW energy density is transferred entirely into DR, 
the extra number of neutrinos species receives a correction \cite{Gouttenoire:2023gbn}
\be
    \Delta N_{\rm eff} = \frac{8}{7}\left(\frac{11}{4}\right)^{4/3}\frac{g_*(T_{\rm BBN})}{2}\alpha_{\rm ann}\,,
\ee
where the effective number of SM relativistic degrees of freedom at the time of BBN is $g_*(T_{\rm BBN}) \simeq 3.36$.
This quantity is constrained by BBN, $\Delta N_{\rm eff} \leq 0.3$ \cite{Fields:2019pfx}, leading to an upper bound on the amplitude of GWs generated by a network of DWs decaying into DR. The bound is shown in pink in Fig.~\ref{fig:DWGW}. On the other hand, if DWs mainly decay into SM degrees of freedom, such an upper bound is relevant only for networks decaying after the neutrino decoupling temperature, $T_{\rm ann} \lesssim 1\,{\rm MeV}$ \cite{Kawasaki:2000en, Bai:2021ibt}. Indeed, in this scenario, neutrinos produced by the DW decay would not be in thermal equilibrium by the time of BBN. Such a low annihilation temperature, however, is associated to much lower frequencies of the GW signal.

\section{Conclusions}

Models with accidentally light scalars \cite{Brummer:2023znr} are particularly well suited for realizing hybrid inflation. They feature tree-level flat directions, lifted by loop corrections, along which slow-roll inflation takes place. They also feature tree-level couplings of the inflaton to other scalar fields, which can become tachyonic along the inflationary direction, thereby ending inflation by the waterfall mechanism.

In this paper, we have presented a simple yet realistic example. 
It is defined by a symmetry group $G=\SO{3}$ and two scalar multiplets $\phi$ and $\chi$, transforming in the representation $\boldsymbol{5}$ and $\boldsymbol{3}$, respectively. The two fields may also carry an abelian charge (discrete or continuous), depending on the variant of the model.
Gauging $\SO{3}$ and allowing for all renormalizable potential terms compatible with the symmetry, one degree of freedom $a$ in $\phi$ remains massless at tree level: This is what we call an
accidentally light field, which we identify with the inflaton. The inflationary potential arises at the one-loop level, induced by the gauge coupling and a quartic coupling between $\phi$ and $\chi$. It has a form similar to that of Natural Inflation \cite{Freese:1990rb}. One component of $\chi$ plays the role of the waterfall field: Its mass-squared starts out positive and decreases during inflation until, at some critical point along the inflationary direction, it turns tachyonic. This destabilizes the waterfall field, which terminates inflation by fast-rolling down its steep potential. 

The flatness of the inflationary potential is naturally protected: It is guaranteed by symmetry at the tree level, and it gets lifted by one-loop radiative corrections. Unlike pseudo-Goldstone inflatons, here the inflaton has tree-level non-derivative couplings to other scalars, which are necessary for an efficient waterfall. These couplings are typically a concern in earlier realizations of hybrid natural inflation, as they spoil the flatness, unless they are forbidden by some complicated additional symmetries or fine-tuned.

The predictions of the model are in agreement with the latest CMB measurements \cite{Planck:2018vyg, Planck:2018jri} for a wide region of the parameter space. We identify a portion of parameter space for which the parameters are natural (assuming the UV cutoff is identified with $\MP$), and fine-tuning of the initial and final points of inflation is less than 10\%. Inside this region, the predicted tensor-to-scalar ratio is $ \left(4\times 10^{-2}\right)n_r\lesssim r \lesssim 4\,n_r$, with $n_r$ natural in the range $6.5\times 10^{-5}\lesssim n_r \lesssim 2\times 10^{-3}$.

Inflation ends as the waterfall field becomes tachyonic, and the Universe is reheated with a reheating temperature of the order of the inflationary scale, $T_{\rm RH}\sim{\cal O}(V_0^{1/4})$. A negative curvature of the potential triggers tachyonic instability of the waterfall field quantum fluctuations, which are exponentially enhanced. This leads to production of GWs, which reach us as a SGWB signal. The peak frequency of the GW spectrum is typically as large as $\mathcal{O}\left(10^{10}\right)\,{\rm Hz}$, but it can be lowered by sufficiently decreasing the inflationary scale. We find that the signal of GWs produced during tachyonic instability is within the reach of the Einstein Telescope for an inflationary scale in the range $3\times 10^5\,{\rm GeV}\lesssim V_0^{1/4}\lesssim 5\times 10^{11}{\rm GeV}$. 
Such a low inflationary scale, despite pointing to some UV completion well below $\MP$, is perfectly viable and is associated to an unobservably small tensor-to-scalar ratio. This confirms that GWs from a tachyonic instability can act as a possible probe of low-scale inflationary scenarios.

The inflationary phase may be followed by a production of topological defects, depending on the specific symmetries which undergo SSB. 
In a model variation with $\phi$ and $\chi$ charged under an additional $U(1)$ gauge symmetry, the vacuum manifold is such that CSs are produced during the waterfall transition. The tension of the string network is bounded by CMB, and this constrains the inflationary scale to be $V_0^{1/4}\lesssim$ a few $\times 10^{14}\,{\rm GeV}$. 
Another model variation features a softly-broken $\mathbb{Z}_4$ symmetry acting on $\phi$ and $\chi$. In this case, the vacuum manifold is disconnected, and unstable DWs are generated at the end of inflation. They produce a SGWB, which can fit the signal recently detected by the NANOGrav Collaboration \cite{NANOGrav:2023hvm}, albeit for an extremely low inflationary scale, $V_0^{1/4} \simeq 2\times 10^{5}\,{\rm GeV}$,  assuming order one quartic couplings.
The DW annihilation signal can also be relevant for future detection by LISA (ET) for $3\times10^7\,{\rm GeV}\lesssim V_0^{1/4}\lesssim 2\times10^{10}\,{\rm GeV}$  ($3\times10^{10}\,{\rm GeV} \lesssim V_0^{1/4}\lesssim 7\times 10^{12}\,{\rm GeV}$). 

Upcoming experiments will soon be able to further constrain or rule out our models, for example, by measuring the sign of $n_r$, or by a discovery of a SGWB by the LIGO-Virgo-KAGRA network. Concerning future directions, it would be interesting to couple our models to the Standard Model in order to study reheating in detail. Finally, the accident mechanism could be applied to other realizations of inflation.

\section*{Acknowledgements}
We thank Pierre Auclair, Yann Gouttenoire, Thomas Hambye, Antonio Junior Iovino, Lucas Pinol, Cristophe Ringeval, Simon Schreyer, and Alfredo Urbano for useful discussions. M.~F.~has received support from the European Union Horizon 2020 research and innovation program under the Marie Sk\l odowska-Curie Grant Agreements No 860881-HIDDeN and No. 101086085–ASYMMETRY.

\appendix 
\counterwithin*{equation}{section}
\renewcommand\theequation{\thesection\arabic{equation}}

\section{Details on the vacuum structure of the model of Sec.~\ref{sec:AccInf}}
\label{sec:AppendixDetails}

In this appendix, we collect some technical details concerning the minimization of the scalar potential in Eq.~\eqref{V5}, as well as the computation of the associated one-loop effective potential.

\subsection{Tree-level analysis}

Eq.~\eqref{V5} is the most general polynomial of degree $\leq 4$ in $\phi$ and $\chi$ which is compatible with $\SO{3}\times\mathbb{Z}_2$ invariance (where $\mathbb{Z}_2$ acts as $\phi\into -\phi$). To see this, note that each term must contain an even number of $\phi$ factors, and that the symmetric products between two $\phi$'s and between two $\chi$'s decompose as $({\mathbf{5}}\otimes {\mathbf{5}})_{\rm sym}={\mathbf{1}}\oplus{\mathbf{5}}\oplus{\mathbf{9}}$ and $({\mathbf{3}}\otimes{\mathbf{3}})_{\rm sym}={\mathbf{1}}\oplus{\mathbf{5}}$, respectively. 
To form a singlet bilinear, the only possibilities are $(\phi\phi)_{\mathbf{1}}$ and $(\chi\chi)_{\mathbf{1}}$. Cubic $\SO{3}$-invariants can be constructed as $(\phi\phi)_{\mathbf{5}}\phi$ and $(\chi\chi)_{\mathbf{5}}\phi$, but these are forbidden by the $\mathbb{Z}_2$ symmetry.  Invariant quartic terms are $(\phi\phi)_\mathbf{1}(\phi\phi)_\mathbf{1}$, $(\phi\phi)_\mathbf{5}(\phi\phi)_\mathbf{5}$ and $(\phi\phi)_\mathbf{9}(\phi\phi)_\mathbf{9}$, the latter two of which can be checked to be proportional to the first; $(\chi\chi)_\mathbf{1}(\chi\chi)_\mathbf{1}$ and $(\chi\chi)_\mathbf{5}(\chi\chi)_\mathbf{5}$, which are likewise proportional to each other; $(\chi\chi)_\mathbf{1}(\phi\phi)_\mathbf{1}$ and $(\chi\chi)_\mathbf{5}(\phi\phi)_\mathbf{5}$. There are, therefore, four independent quartic invariants, which may be represented by the four quartic terms in Eq.~\eqref{V5}.

Moreover, the largest continuous symmetry under which Eq.~\eqref{V5} is invariant is indeed $\SO{3}$. We have checked this explicitly, by considering the full $\SO{8}$ symmetry which acts on the $\phi$ and $\chi$ components in the absence of any potential.
We verified that only three $\SO{8}$ generators are preserved by $V$, and they form an $\SO{3}$ algebra. This is a nontrivial statement; cf., e.g., the electroweak sector of the Standard Model, where the most general renormalizable potential for the Higgs doublet which is invariant under $\SU{2}\times\U{1}$ gauge transformations turns out to preserve a larger, ``custodial'' $\SO{4}$ symmetry.

To find the extrema of the tree-level potential of Eq.~\eqref{V5}, we will limit ourselves to the case of positive quartic couplings, which is the one of interest for our inflationary model. For definiteness, we use the $5$-plet generators
\be
\begin{array}{c}
T^1={\scriptsize{\left(\begin{array}{ccccc}&&&-i& \\ &&&&-i \\ &&&\sqrt{3} i& \\ i&&-\sqrt{3} i&& \\ &i&&&\end{array}\right)}},\,
T^2={\scriptsize{\left(\begin{array}{ccccc}&-i&&& \\ i&&\sqrt{3}i&& \\ &-\sqrt{3}i&&& \\ &&&&i \\ &&&-i&\end{array}\right)}},\,
T^3={\scriptsize{\left(\begin{array}{ccccc}&&&&-2i \\ &&&-i& \\ &&0&& \\ &i&&& \\ 2i&&&&\end{array}\right)}}\,.
\end{array}\ee
The critical points are most easily identified after using
$\SO{3}$ invariance to rotate $\chi\into (0,\, 0,\, \chi_3)$. The potential becomes
\be
V=-\frac{1}{2}\mu_\phi^2\phi^2-\frac{1}{2}\mu_\chi^2\,\chi_3^2+\frac{\lambda_\phi}{4}(\phi^2)^2+\frac{\lambda_\chi}{4}\chi_3^4+\frac{\varepsilon}{4}\phi^2\chi_3^2+
 \frac{\zeta}{4}\chi_3^2\,\phi^{\rm T}(T^3)^2\phi\,,
\ee
with $(T^3)^2={\rm diag}\,(4,1,0,1,4)$. Hence, the quartic terms are all positive definite, which shows, in particular, that the potential is bounded from below. We have
\be
\begin{split}
\frac{\partial V}{\partial\phi_A}=&\;
\left[-\mu_\phi^2 +\lambda_\phi\,\phi^2+\frac{\varepsilon}{2}\chi_3^2+\frac{\zeta}{2}\chi_3^2(T^3)^2_{AA}\right]
\phi_A\,,\\
\frac{\partial V}{\partial\chi_3}=&\;\left[-\mu_\chi^2+\lambda_\chi\,\chi_3^2+\frac{\varepsilon}{2}\phi^2
+\frac{\zeta}{2}\phi^{\rm T}(T^3)^2\phi\right]\chi_3\,.
\end{split}
\ee
The critical points are, thus, seen to be as follows.
\begin{itemize}

 \item $\phi=0$, $\chi=0$.---This is a local minimum if $\mu_\phi^2<0$ and $\mu_\chi^2<0$, a local maximum if both are positive (which is the case in our inflationary model), and a saddle point otherwise.
 
 \item $\chi=0$ and  $\phi^2=\mu_\phi^2/\lambda_\phi\equiv v^2$
 (assuming $\mu_\phi^2>0$).--- Therefore, this set of critical points forms a four-sphere. 
 The mass matrix does not mix $\phi$ and $\chi$, and the $\phi$ block is easily seen to have eigenvalues $(2\lambda_\phi v^2,0,0,0,0)$. For a generic point on the vacuum manifold, there are three Nambu-Goldstone directions, as expected for a completely broken $\SO{3}$ symmetry, and one additional, accidentally flat
 direction. However, the ${\bf 5}$-plet generators each have a zero eigenvalue, and, therefore, annihilate a $\phi$ VEV oriented in the direction of the corresponding eigenvector. These special points in the vacuum manifold preserve an O$(2)\subset\SO{3}$ symmetry, and, thus, feature two Nambu-Goldstone fields and two accidentally massless fields. All such points are related to one another by $\SO{3}$ transformations, and, therefore, correspond to a single physical point when $\SO{3}$ is gauged.
 
 The $\chi$ block of the mass matrix is 
 \be
 \left.\frac{\partial^2 V}{\partial\chi_a\partial\chi_b}\right|_{\chi=0,\,\phi^2=v^2}=\left(\frac{\varepsilon}{2}v^2-\mu_\chi^2\right)\delta^{ab}+\frac{\zeta}{2}\phi_A T^a_{AC} T^b_{CB}\phi_B\,.
 \ee
The last term is diagonalized by an $\SO{3}$ transformation which sends $\phi_{2,4,5}\into 0$, with eigenvalues $\frac{\zeta}{2}(\phi_1\pm\sqrt{3}\phi_3)^2$ and $2\zeta\phi_1^2$. The $\chi$ masses are, therefore, positive for $\mu_\chi^2<0$, but for $\mu_\chi^2>0$ the sign of the mass terms depends on the direction of the $\phi$ VEV, as detailed in the main text. This is the field configuration realized on the inflationary trajectory.

\item $\phi=0$ and $\chi^2 = \mu_\chi^2/\lambda_\chi \equiv v_\chi^2 $ (assuming $\mu_\chi^2>0$).---The mass matrix in the $\chi$ sector has eigenvalues $(2\lambda_\chi v_\chi^2,0,0)$, with two Nambu-Goldstone directions as expected for $\SO{3}\into\SO{2}$ breaking by a triplet VEV. The mass matrix in the $\phi$ sector is diagonal in a basis where $\chi_{1,2}=0$ and $\chi_3=v_\chi$:
 \be
  \left.\frac{\partial^2 V}{\partial\phi_A\partial\phi_B}\right|_{\chi=\begin{psmallmatrix} 0\\0\\v_\chi\end{psmallmatrix},\, \phi=0}=\bigg[\frac{\varepsilon}{2}v_\chi^2-\mu_\phi^2+\frac{\zeta}{2}v_\chi^2(T^3)^2_{AA}\bigg]\delta_{AB}\,.
 \ee
 Therefore, this critical point is a minimum provided that $\frac{\varepsilon}{2}v_\chi^2>\mu_\phi^2$. It is the true vacuum of our inflationary model, by a choice of parameter values.
 
 \item There are also critical points with both $\phi$ and $\chi$ nonzero. In a basis where $\chi=(0,\,0,\,\chi_3)$, one finds that this requires either $\phi_{1,2,4,5}=0$ or $\phi_{1,3,5}=0$ or $\phi_{2,3,4}=0$. Explicit expressions for the expectation values of the other fields can be derived, but we refrain from giving them here, since these critical points are saddle points for the parameter choices of interest.
\end{itemize}

\subsection{One-loop corrections}

At the one-loop level, the tree-level scalar potential is corrected by the well-known Coleman-Weinberg effective potential
\be
V_{\rm eff}=\frac{1}{64\pi^2}\,\tr\,\left({\cal M}^4\log\frac{{\cal M}^2}{\Lambda^2}\right)\,.
\ee
Here, ${\cal M}$ is the scalar-field-dependent mass matrix for gauge fields and scalars (with a factor 3 for the former, accounting for three polarizations for a massive vector field) and $\Lambda$ is the renormalization scale. 

We are interested, notably, in the loop-corrected effective potential for a field configuration with VEVs $\phi^2=v^2$ and $\chi=0$, along the direction which remained accidentally flat at the tree level. This direction was parameterized by an angle $\alpha\in[0,\,\pi/6]$ in the main text; see Eq.~(\ref{eq:alpha}). Evaluating $V_{\rm eff}$ at some fixed value of $\alpha$, $\bar{\alpha}=\pi/6$, say, we find that there is a divergent tadpole term for the radial mode $\rho$, as well as a divergent contribution to the vacuum energy density. These divergences need to be subtracted by including counterterms for all parameters of the model and imposing suitable renormalization conditions. A particularly convenient set of conditions is to (i) renormalize the $\rho$ tadpole to be zero, such that $v^2\equiv m_\phi^2/\lambda_\phi$  remains the VEV after including one-loop corrections (this is an implicit condition on the counterterms for $m^2_\phi$ and $\lambda_\phi$); and (ii) set the renormalized vacuum energy density to some finite constant. There are other divergences in the one-loop effective action, but these two conditions suffice to render the effective potential along the $\alpha$ direction finite. It is given by \cite{Brummer:2023znr}
\be\begin{array}{l}
V_{\rm eff}(\alpha)=\dfrac{1}{64\pi^2}\times
\\
\times\Biggl[\tr\left( M^4(\alpha)\log \dfrac{M^2(\alpha)}{\Lambda^2}-M^4(\bar{\alpha})\log \dfrac{M^2(\bar{\alpha})}{\Lambda^2}\right)\\
+3\,\tr \left( M_W^4(\alpha)\log \dfrac{M_W^2(\alpha)}{\Lambda^2}- M_W^4(\bar{\alpha})\log \dfrac{M_W^2(\bar{\alpha})}{\Lambda^2}\right)\Biggr]
\,,
\end{array}\ee
where $M$ and $M_W$ are the scalar and gauge boson mass matrices, respectively, and the $\Lambda$ dependence is, in fact, spurious and cancels out, $\tr M^4$ and $\tr M_W^4$ being $\alpha$-independent. Note that the choice of the subtraction point $\alpha=\bar{\alpha}=\pi/6$ is arbitrary: We could have chosen any other point provided that the tree-level mass matrix is positive definite, so that the effective potential is well-defined. The gauge symmetry leads to the identifications $\alpha\into -\alpha$ and $\alpha\into\alpha+\pi/3$; therefore, the effective potential is even and periodic. As such, $V_{\rm eff}(\alpha)$ can be expanded into a Fourier cosine series, $V_{\rm eff}(\alpha)=\sum_n c_n \cos(6n\alpha)$. The leading Fourier mode of $V_{\rm eff}$ is then found to be given by the expressions of Eqs.~\eqref{eq:CosPot} and \eqref{eq:M4}, and it provides a slow-roll potential for the inflaton field.

\section{A model with complex scalars, giving rise to cosmic strings}
\label{sec:AppendixComplexModel}

In Sec.~\ref{sec:AccInf}, we presented a minimal model with two real scalar $\SU{2}$ multiplets, a five-plet $\phi$ and a triplet $\chi$. Here, we take these fields to be complex, charged under additional \U{1} symmetries.

Let $\phi$ be a complex scalar transforming in the {\bf 5} of \SU{2} with unit $\U{1}_{\phi}$ charge. The most general renormalizable potential is
\be\label{V5Compl}
 V_\phi=-\mu_{\phi}^2\,\phi^\dag\phi+\lambda \,(\phi^\dag\phi)^2+ \kappa
 \left[(\phi^\dag\phi)^2-|\phi^T\phi|^2\right]+ \delta\, (\phi^\dag T^a\phi)^2\,,
\ee
with the generators $T^a$ defined below Eq.~\eqref{V5}.
A detailed study of this potential can be found in Ref.~\cite{Brummer:2023znr}.\footnote{The quartic couplings $\lambda,\kappa$, and $\delta$ here are normalized to be a factor 2 smaller than in Ref.~\cite{Brummer:2023znr}.} The quartic couplings $\lambda$, $\kappa$, and $\delta$ and the mass parameter $\mu_\phi^2$ are chosen positive. Notice that the $\kappa$ and $\delta$ terms vanish when $\phi$ is real.
Cubic terms are forbidden by $\U{1}_{\phi}$, with no need for ad-hoc discrete symmetries. The minimum of the potential is at $\vev{\phi^\dag\phi}=v^2/2$, where $v^2=\mu_\phi^2/\lambda$. When $\SU{2}\times\U{1}_\phi$ is gauged, all NGBs are absorbed by the gauge bosons. However, there remains a continuous family of vacua at the tree level, parameterized by an angle $\alpha\in[0,\pi/6]$ as in Eq.~\eqref{eq:alpha}. 
Thus, $\alpha$ is an accidentally flat direction. For generic $\alpha$, all of the gauge symmetry is broken, but, at the special point $\alpha=0$, a subgroup $\U{1}'\subset\SU{2}$ remains unbroken. At this point, there is an additional massless scalar in the spectrum, besides the one corresponding to the flat direction.

In the limit $\delta \rightarrow 0$, the potential recovers an enlarged $\SO{5}\times \U{1}_{\phi}$ symmetry, spontaneously broken to $\SO{4}$. Switching on $\delta$ explicitly breaks the symmetry of the potential to $\SU{2}\times \U{1}_{\phi}$, so that the continuous symmetry of $V$ is strictly no larger than the gauge symmetry: Therefore, the accidentally light mode is not a pNGB in the sense of \cite{Weinberg:1972fn}.

Loop corrections lift the accidentally flat direction, analogously to the minimal model in Sec.~\ref{sec:AccInf}, with $\alpha=0$ being selected as the true vacuum. The one-loop effective potential along the tree-level flat direction takes the form $V_{\rm eff}\simeq V_0' -\left(M'\right)^4 \cos \left(a/f\right)$ (up to higher harmonics whose coefficients are numerically small for all values of the couplings of interest). The amplitude of the potential is given by 
\be\label{eq:M4Pr}
\left(M'\right)^4 =\frac{v^4}{160\,\pi^2}\Bigg[9\,g^4-\frac{\kappa^5}{\delta^3} \Bigg(F\left(-\frac{2\delta}{\kappa}\right)-T_{F,5}\left(-\frac{2\delta}{\kappa}\right)\Bigg)\Bigg]\,,
\ee
where $g_2$ is the $\SU{2}$ gauge coupling, and $F(x)$ and $T_{F,5}$ are defined below Eq.~\eqref{eq:M4}.
The tree-level massless scalar picks up a mass which is one-loop suppressed with respect to the masses of the other fields.

Now add to the model a complex scalar $\chi$ transforming in the ${\bf 3}$ of $\SU{2}$ with unit $\U{1}_\chi$ charge, which will play the role of the waterfall field. 
The most general renormalizable scalar potential now also includes the terms
\be
\begin{aligned}
V_\chi=&\;-\mu^2_\chi\chi^\dag\chi
+\lambda_\chi\,(\chi^\dag\chi)^2+\lambda'_\chi\,|\chi^T\chi|^2\,,
\\
V_{\phi\chi}=& \;\varepsilon\,\phi^\dag\phi\chi^\dag\chi 
+\zeta\,T^a_{AB} T^b_{BC}\phi^*_A \phi_C \chi^{*a}\chi^b+\vartheta\,T^a_{AB}(i\varepsilon^{abc})\phi^*_A \phi_B\chi^{*b}\chi^c\,.
\label{Vchi}
\end{aligned}
\ee
Notice that one can replace
the phase rotations $\U{1}_{\phi}\times \U{1}_\chi$ by a single gauged $\U{1}$, with charges $q_\phi$ and $q_\chi$ chosen to forbid additional mixed couplings, beside $\epsilon,\zeta$, and $\vartheta$.
In this case, by construction, the scalar potential has an accidental global $\U{1}_{\phi}\times\U{1}_\chi$
symmetry. As a consequence, when $\langle \chi \rangle \ne 0$ is generated after inflation, one produces cosmic strings with observable consequences, as discussed in Sec.~\ref{sec:TopologicalDefects}.

For $\mu^2_\chi<0$ and positive quartic couplings in Eq.~\eqref{Vchi}, the tree-level potential is easily seen to be minimized at $\chi=0$, with the tree-level vacuum structure for $\phi$ unchanged. In fact, when $\phi$ is set to its tree-level minimum of Eq.~\eqref{eq:alpha}, the terms $\mu^2_\chi$ and $\varepsilon$ give rise to a universal, $\alpha$-independent mass term for the three components of $\chi$, while the $\zeta$ term splits their masses in an $\alpha$-dependent fashion (and the $\vartheta$ term does not contribute to the mass matrix). 
The mass eigenvalues for the $\chi$ fields are the same as in Eq.~\eqref{eq:mchi}.
To obtain a model of hybrid inflation, we choose $\tilde\mu_\chi^2\equiv\mu_\chi^2-\frac{\varepsilon}{2}v^2>0$, and $\zeta v^2>2\tilde\mu_\chi^2$, such that the overall $\chi^3$ mass-squared is positive for $\alpha=\pi/6$ but negative for $\alpha=0$. At the intermediate value $\alpha=\alpha_c$ (given in Eq.~\eqref{eq:alphac}),
the effective $\chi^3$ mass crosses zero.

For suitable values of the couplings, the true vacuum of the model is located at
\be
\langle\lvert\chi^3\rvert\rangle=\sqrt{\frac{\mu^2_\chi}{2\left(\lambda_\chi+\lambda_\chi'\right)}}\,,\qquad
\text{all other fields} = 0
\ee
up to gauge transformations. The residual gauge symmetry is $\U{1}'\subset \SU{2}$, and all fields except the unbroken gauge boson pick up tree-level masses.
This holds for $\lambda_\chi'>0$, while for $-\lambda_\chi <\lambda_\chi'<0$, the VEV of $\chi$ is aligned differently, and it breaks $\SU{2}$ entirely.

Loops of $\chi$ contribute to the accident effective potential so that Eq.~\eqref{eq:M4Pr} becomes
\be\label{eq:M4Przeta}
\begin{split}
\left(M'\right)^4 =\frac{v^4}{160\,\pi^2}\Bigg[&9\,g_2^4-\frac{\kappa^5}{\delta^3}\left(F\left(-\frac{2\delta}{\kappa}\right)-T_{F,5}\left(-\frac{2\delta}{\kappa}\right)\right)\\&+4\,\frac{\tilde\mu_\chi^{10}}{\zeta^3\,v^{10}}\left({\rm Re}\;F\left(\frac{\zeta v^2}{\tilde\mu_\chi^2}\right)-T_{F,5}\left(\frac{\zeta v^2}{\tilde\mu_\chi^2}\right)\right)\Bigg]\,.
\end{split}
\ee
Note that the ${\rm Re}\,F$ term in the second line vanishes for the parameter region of physical interest $\tilde\mu_\chi^2>0$.

The calculation of CMB observables in this model is done in the same way as outlined in Sec.~\ref{sec:CMB}, upon replacing $M\into M'$ and $V_0\into V_0'$.

\section{A model with disconnected minima, giving rise to unstable domain walls}
\label{sec:AppendixZ4Model}

In this variant of the model, the potential has two physically distinct minima which can be chosen to be near-degenerate.
Let $\phi$ be a real five-plet of $\SO{3}$ which is odd under $\mathbb{Z}_{2\phi}:\phi\to-\phi$, and add to the model a pair of triplet waterfall fields $\chi_R$ and $\chi_I$; we represent them as a single complex field $\chi=(\chi_R+i\chi_I)/\sqrt{2}$, although no $\U{1}$ symmetry is implied. We do, however, impose a $\mathbb{Z}_{4\chi}$ symmetry acting as $\chi \rightarrow i \chi$. 
With these fields and symmetries, the most general renormalizable potential reads
\be
\begin{aligned}
 V=&-\frac{1}{2}\mu_\phi^2\phi^2+\frac{\lambda_\phi}{4}(\phi^2)^2-\mu_\chi^2\chi^* \chi +\lambda_\chi\left(\chi^*\chi\right)^2+\delta\,{\chi^*}^2\chi^2
 + \dfrac 12\left(\kappa\, \chi^2 \chi^2+{\rm h.c.}\right)\\
 &+\frac{\varepsilon}{2}\phi^2\left(\chi^*\chi\right)+
 \frac{\zeta}{2}\,T^a_{AC} T^b_{CB}\phi_{A} \phi_B {\chi^{ a}}^*\chi^{ b}\,,
\end{aligned}
\label{VZ4}\ee
where $T^a$ are the $\SO{3}$ five-plet generators, chosen as $5 \times 5$ imaginary and anti-symmetric matrices. For simplicity, we assume $\kappa$ real, which corresponds to a CP invariance $\chi\leftrightarrow\chi^*$, and all the quartic couplings to be positive. 

Let us first consider the case $\mu_\phi^2>0$ and $\mu_\chi^2<0$. The minimum is at $\langle \phi^2 \rangle =v^2\equiv\mu_\phi^2/\lambda_\phi$, $\langle \chi \rangle=0$, where the mass spectrum resembles the one of the minimal model of Sec.~\ref{sec:AccInf}. After the spontaneous breaking of $\SO{3}$, one observes the emergence of an accidentally flat direction parametrized by Eq.~\eqref{eq:alpha}. The masses of the complex $\chi$ components are given by Eq.~\eqref{eq:mchi}. 
One-loop corrections give rise to the usual potential along the accidentally flat direction $V_{\rm eff} = V_0'' - \left(M''\right)^4\cos\left(a/f \right)$, with $M''$ given by
\be
\left(M''\right)^4 
=\frac{1}{160\,\pi^2}\Bigg[9\,g^4\,v^4+4\,\frac{\tilde\mu_\chi^{10}}{\zeta^3\,v^6}
\Bigg({\rm Re}\;F\left(\frac{\zeta v^2}{\tilde\mu_\chi^2}\right)-T_{F,5}\left(\frac{\zeta v^2}{\tilde\mu_\chi^2}\right)\Bigg)\Bigg]\,,
\ee
and $F$ and $T_{F,5}$ defined below Eq.~\eqref{eq:M4}.

We now switch to the case interesting for inflation: $\mu_{\phi}^2>0$ and $\mu_{\chi}^2>0$. The $\chi_3$ mass is positive at the beginning of inflation, close to $\alpha = \pi/6$, and it crosses $0$ at $\alpha = \alpha_c$, given by Eq.~\eqref{eq:alphac}.
With this choice of couplings, the potential has a minimum at $\langle\phi\rangle=0$ and $\langle\chi\rangle\neq0$. The six equations to solve in order to find the extrema of the potential are 
\be
\begin{aligned}
    \chi_R^a\left[-\mu_{\chi}^2 + \lambda_\chi \left(\chi_R^2+\chi_I^2 \right)+(\kappa + \delta)(\chi_R^2 -\chi_I^2)\right]-2\chi_I^a \left(\chi_R\cdot\chi_I\right)\left(\kappa-\delta\right)&=0\,,\\
   \chi_I^a\left[-\mu_{\chi}^2 + \lambda_\chi \left(\chi_R^2+\chi_I^2 \right)-(\kappa + \delta)(\chi_R^2 -\chi_I^2)\right]-2\chi_R^a \left(\chi_R\cdot\chi_I\right)\left(\kappa-\delta\right)&=0\,,
\end{aligned}
\ee
for $a=1,2,3$. It is convenient to employ $\SO{3}$ invariance to set $\chi^1_R =\chi^2_R=0$ as well as 
$\chi^1_I=0$. Then, the system reduces to
\be
\begin{aligned}
    &\chi^2_I \chi^3_I \chi^3_R =0
    \,,\\
    &\chi_{I}^2\left[-\mu_\chi^2+\lambda_\chi
    [(\chi_R^3)^2+(\chi_I^2)^2+(\chi_I^3)^2]
    -(\kappa+\delta)
    [(\chi_R^3)^2-(\chi_I^2)^2-(\chi_I^3)^2]\right]=0
    \,,\\
    &\chi_{R}^3\left[-\mu_\chi^2+\lambda_\chi
    [(\chi_R^3)^2+(\chi_I^2)^2+(\chi_I^3)^2]
    +(\kappa+\delta)
    [(\chi_R^3)^2-(\chi_I^2)^2]
    - (3\kappa-\delta)(\chi_{I}^3)^2\right]=0
    \,,\\
     &\chi_{I}^3\left[-\mu_\chi^2+\lambda_\chi
     [(\chi_R^3)^2+(\chi_I^2)^2+(\chi_I^3)^2]
     +(\kappa+\delta)
     [(\chi_I^2)^2+(\chi_I^3)^2]
     - (3\kappa-\delta)(\chi_{R}^3)^2\right]=0\,.
\end{aligned}
\ee

For a certain hierarchy among the quartic couplings
($0<\delta<\kappa<\lambda_\chi+\delta < \varepsilon/2+\kappa$), the potential is minimized at
\be\label{eq:appBvacuum}
(\chi_R^3)^2 =(\chi_I^3)^2 = w^2 \equiv \frac{\mu_{\chi}^2}{2
\left(\lambda_\chi+\delta-\kappa\right)}\,,
\quad \chi_I^2=0\,.
\ee
In these minima, the symmetry is spontaneously broken as $\SO{3}\times \mathbb{Z}_{4\chi}
\to \SO{2}\times \mathbb{Z}_{2\chi}$. 
There are four solutions to Eq.~\eqref{eq:appBvacuum} but only two physically distinct vacua: one $(+)$ where the $\chi_R$ and $\chi_I$ VEVs are parallel, $\chi_R^3 = \chi_I^3 = \pm w$, and the other $(-)$ where they are antiparallel, $\chi_R^3 =-\chi_I^3 = \pm w$. 
Thus, the vacuum manifold has two disconnected components, and DWs can be produced during the SSB phase transition, as discussed in Sec.~\ref{sec:TopologicalDefects}.

For a viable phenomenology, the DW network needs to be unstable, as discussed in Sec.~\ref{secDW}. We, therefore, assume that $\mathbb{Z}_{4\chi}$ is softly broken by a term $V_{\rm soft}=i m_\chi^2 \chi\chi + {\rm h.c.}$, with $m_\chi^2$ real. This term preserves $\mathbb{Z}_{2\chi}:\chi\to-\chi$, while it breaks the CP symmetry $\chi\to\chi^*$. Consequently, it breaks the degeneracy between the two disconnected vacua ($+$) and ($-$), introducing a bias in the potential. We assume that soft $\mathbb{Z}_{4\chi}$ breaking is a small effect,  $m_\chi^2 \ll \mu_{\chi}^2$. Then the potential bias is, to leading order in $m_\chi^2$,
\be\label{eq:Vbias}
 \Delta V \equiv V_{(+)} - V_{(-)}= - \frac{2\,m_\chi ^2\mu_{\chi}^2}{\lambda_\chi+\delta-\kappa} + \mathcal{O}\left(m_\chi^4\right)\,,
\ee
where $V_{(+)}$ and $V_{(-)}$ are the potential energies evaluated at the minimum configuration in which $\chi_R$ and $\chi_I$ are parallel and anti-parallel, respectively.

\bibliographystyle{hieeetr}
\bibliography{AccInf.bib}

\end{document}